\documentclass[12pt,a4paper]{article}

\usepackage{amsmath,amssymb,bm,graphicx}

\begin{document}

\begin{center}
  \vspace*{\fill}
  \vspace*{\fill}
  {\Large
    Density Functional Application to\\
    Strongly Correlated Electron Systems\\
  }
  \vspace*{\fill}\vspace*{\fill}
  \begin{minipage}{6cm}
    H. Eschrig\\
    K. Koepernik\\
    I. Chaplygin
  \end{minipage}\\
  \vspace*{\fill}
  \begin{minipage}[t]{6cm}
    email: h.eschrig@ifw-dresden.de\\
    Tel: +49(0)351-4659-380\\
    Fax: +49(0)351-4659-500\\
  \end{minipage}
  \vspace{\fill}

  \textbf{This paper is dedicated to Walter Kohn on the occassion of his\\
    eightieth birthday in March 2003}
\end{center}
  \vspace*{\fill}
  \vspace*{\fill}
  \vspace*{\fill}

\newpage

\abstract{ The LSDA+$U$ approach to density functional theory is carefully
  reanalyzed. Its possible link to single-particle Green's function theory is
  occasionally discussed. A simple and elegant derivation of the important
  sum rules for the on-site interaction matrix elements linking them to the
  values of $U$ and $J$ is presented. All necessary expressions for an
  implementation of LSDA+$U$ into a non-orthogonal basis solver for the
  Kohn-Sham equations are given, and implementation into the FPLO solver
  \cite{Koe99} is made. Results of application to several planar cuprate
  structures are reported in detail and conclusions on the interpretation of
  the physics of the electronic structure of the cuprates are drawn.
}

\newpage

\section{Introduction}

Density functional theory (DFT) based on the variational principle by
Hohenberg and Kohn \cite{Hoh64} has nowadays a rigorous mathematical basis,
mainly due to work by Lieb \cite{Lie83}. As a theory for (chosen) ground state
properties of a many-particle system, it holds true for any Coulomb quantum
system with arbitrarily strong correlations in the ground state. All hampering
representability problems of the early time are gone (see for instance
\cite{Esc96}). However, the central quantity, the universal density functional
is not known, only its existence can be proved, and we have no fully
systematic access by approximations. Hence, so far (and very likely also in
future) we have to model it and to probe the models by comparison to
phenomenology. This situation is not principally different from other
many-particle approaches where either models of sufficiently simple
Hamiltonians are used (in quantum field theory) or the wave function is
modeled (for instance in Hartree-Fock or Gutzwiller approaches).

The situation is even less satisfactory in solid state theory, if the focus is
on the excitation spectra instead on the ground state, because in most cases
the spectrum of the many-body Hamiltonian has no separate physical relevance
at all except for its formal use in theoretical expressions for the partition
function. Instead, what is measured are the spectra of various quasistationary
excitations, defined from few-particle Green's functions, the self-energy
parts of which, besides being energy dependent non-linear integral operators,
are also density functionals.

Although this is not a principal restriction of DFT, the models in use so far
(local (spin) density approximation, L(S)DA, in the following the acronym LSDA
is used for both LDA and LSDA, generalized gradient approximation, GGA, LSDA
plus self-interaction correction, SIC, LSDA plus onsite Coulomb repulsion,
LSDA+$U$, \ldots) are subject to the adiabatic approximation for the
electron-lattice interaction.

DFT in the Kohn-Sham (KS) approach to solids yields a KS band structure, which
as such does also not have a direct physical meaning. Instead, the
quasi-particle band structure of Bloch electrons is obtained from the
self-energy of the electron Green's function. It has become common use to
speak of weak correlations, if in the vicinity of the Fermi level the LSDA KS
potential and the electron self-energy are not very different. This does by no
means imply that the correlation energy itself, defined as the difference
between the true total energy and the Hartree-Fock energy, is small or much
smaller than in strongly correlated systems (where the LSDA KS potential
differs strongly from the electron self-energy, the latter often jumps as a
function of energy at the Fermi level). It is a general experience that in the
latter cases also the ground state properties, or certain ground state
properties, are much worse reproduced by the LSDA than in weakly correlated
cases (in the above definition).

This paper deals with strongly correlated systems treated by means of the
LSDA+$U$ approach. Maybe the first precursor of an LSDA+$U$ calculation (at
that time not fully self-consistent) was the treatment of $4f$-electrons by
Herbst \textit{et al.} \cite{Her78}. A first fully self-consistent calculation
of values of the Hubbard $U$ was performed by Dederichs \textit{et al.}
\cite{Ded84} for the $4f$-states of Ce by applying the Koringa-Kohn-Rostoker
solver of the KS equations to a constraint impurity problem. An early similar
approach to $U$ for NiO by Norman and Freeman \cite{Nor86} used the APW solver
in a super-cell formulation. For the further development of this subject see
\cite{Gun89}. 

Pickett and Wang \cite{Pic84} and Hybertsen and Louie \cite{Hyb85} based the
so-called $GW$-approximation for the electron self-energy of semi-conductors
and insulators on LDA results for the density and KS bands as a starting
approximation for the $GW$-approach. In these cases the main difference
between the LDA KS potential and the self-energy is a jump of the latter,
constant in $\bm r$-space at the Fermi level (scissors operation). This is
still considered a weakly correlated case, but these were the first estimates
of the self-energy of an inhomogeneous situation as a density functional.

A systematic incorporation of the Hubbard $U$-potential into the DFT model
functionals started with two papers by Anisimov \text{at al.} \cite{Ani91,
  Ani93}. While it was proposed in \cite{Ani91} to model the total
spin-dependence by the $U$-functional and to treat the spin independent
functional by the LDA, in \cite{Ani93} the orbital polarization part
($m$-dependent occupation of local orbitals) was treated by the $U$-functional
and the isotropic (in $\bm r$-space) part of the spin-density was treated in
LSDA. This has the advantage that spin polarization effects can be treated
more generally, not only in the strongly correlated orbitals. However, this
version, later on called `around the mean field', AMF, by Czy\.zyk and
Sawatzky \cite{Czy94}, gives nearly nothing for a half-filled fully spin
polarized shell as in Mn$^{2+}$ or in Gd. Therefore, aiming mainly at
reproducing the photoemission spectra (which essentially means modeling the
electron self-energy rather that the KS potential), an alternative
$U$-functional was introduced in \cite{Czy94} and called the `atomic limit'
version, AL. This version which roughly shifts unoccupied orbital energies
upward by $U/2$ and occupied orbital energies downward by $U/2$ independent of
the shell filling (even for filled and empty shells), has been widely used
since. For a survey see \cite{Ani97}.

It has to be confessed that all LSDA+$U$ models up to now depend on the basis
used for the KS solver. Most results are obtained so far with an LMTO
implementation. For a recent APW implementation see \cite{Shi99}.

In this paper, in Section 2, an FPLO implementation is described. FPLO
\cite{Koe99} is a high precision high efficiency KS solver which uses a
minimum basis (and hence is fast) containing only local basis functions which
are optimized in both a numerical and chemical sense. (It competes in accuracy
with well converged full-potential APW.) After a short outline how $U$ is
integrated into DFT, the correlated orbitals used in the FPLO implementation
are introduced. Since literature statements \cite{Czy94} say that the most
important sum rules for the interaction matrix elements (screened Slater
integrals) of those orbitals are cumbersome to verify, a very simple and
elegant derivation is given here. After the necessary analysis of the orbital
occupation matrix for the non-orthogonal basis of FPLO and the introduction of
the AMF and AL functionals in the FPLO implementation, explicit expressions
for the $U$-potential and for the total energy are given as they are coded in
FPLO LSDA+$U$. In Section 3, new applications to cuprate structures, the
`infinite layer' compound CaCuO$_2$, the undoped single-layer compound
Sr$_2$CuO$_2$Cl$_2$ and the bilayer high-temperature superconductor
Bi$_2$Sr$_2$CaCu$_2$O$_8$ are considered and the results are compared to both
magnetic ground state properties and photoemission spectra. A short summary is
given in Section 4.

\section{The FPLO implementation of the LSDA+$U$ approach}

The underlying frame of the LSDA+$U$ approach is the Hohenberg-Kohn
variational principle,
\begin{equation}
  \label{eq:1}
  E[\check v, N] = \min_{\check n} \Biggl\{H[\check n] + \sum_{ss'}\int d^3r
  v_{ss'}(\bm r)n_{s's}(\bm r) \;\Biggr|\; \sum_s\int d^3r n_{ss}(\bm r) = N
  \Biggr\} 
\end{equation}
for the ground state energy $E$ and spin density $\check n = (n_{ss'})$ of $N$
electrons in an external spin dependent potential $\check v$,
\begin{equation}
  \label{eq:2}
  \sum_{ss'}\int d^3r v_{ss'}(\bm r)n_{s's}(\bm r) = 
  \int d^3r(vn - \bm{B\cdot m}),
\end{equation}
which holds true in any case of arbitrarily strong correlation. It is based on
many-particle quantum theory by rigorous mathematics. Of course, the density
functional $H[\check n]$ is unknown.

The generalized Kohn-Sham modeling of this functional is by parameterizing the
variational spin density by new variational parameters: the Kohn-Sham orbitals
$\phi_i(\bm r,s)$ and orbital occupation numbers $n_i$,
\begin{equation}
  \label{eq:3}
  n_{ss'}(\bm r) = \sum_i \phi_i(\bm rs)\,n_i\,\phi^*_i(\bm rs'), \;\;
  \langle\phi_i|\phi_j\rangle = \delta_{ij}, \;\;
  0\leq n_i\leq 1, \;\;
  \sum_i n_i = N,
\end{equation}
and by splitting the density functional into an orbital variation expression
$K[\check n]$ and a (possibly generalized by gradient terms) local density
expression $L[\check n]$:
\begin{equation}
  \label{eq:4}
  \begin{split}
      H[\check n] & = K[\check n] + L[\check n] \\
      K[\check n] & = \min_{\{\phi_i,n_i\}} \Bigl\{ k[\phi_i,n_i] \Bigr|
      \sum_i \phi_i n_i \phi^*_i = \check n \Bigr\} \\
      L[\check n] & = \int d^3r\; n(\bm r)\; 
        l\bigl(n_{ss'}(\bm r), \bm\nabla n, \ldots\bigr).
  \end{split}
\end{equation}
This puts the Hohenberg-Kohn variational principle into the Kohn-Sham form
\begin{multline}
  \label{eq:5}
  E[\check v,N] = \min_{\phi_i,n_i} \biggl\{ k[\phi_i,n_i] +
  L\Bigl[\sum_i\phi_in_i\phi^*_i\Bigr] + 
  \sum_i n_i\langle\phi_i|\check v|\phi_i\rangle \;\biggr| \\
  \biggl|\; \langle\phi_i|\phi_j\rangle = \delta_{ij}, \; 0\leq n_i\leq 1, \;
  \sum_i n_i = N \biggr\}.
\end{multline}
While for the $L$-functional LSDA or GGA models are in use, $k$ is modeled
by LSDA or LSDA+SIC or LSDA+$U$. Variation of $\phi_i^*$ yields the
generalized Kohn-Sham equation,
\begin{equation}
  \label{eq:6}
  \biggl[ \frac{1}{n_i}\,\frac{\delta k}{\delta\phi^*} + \check v^L + \check v
  \biggr]\phi_i = \phi_i\varepsilon_i, \quad v^L_{ss'} = 
  \frac{\delta L}{\delta n_{s's}},
\end{equation}
and variation of the $n_i$ yields the common aufbau principle which holds true
for all model variants within this frame and which says that the ground state
density is obtained by occupying the $N$ orbitals with the lowest
$\varepsilon_i$. 

The variants of the LSDA+$U$ model correspond to 
\begin{equation}
  \label{eq:7}
  k = t + e^{\text{H}} + e^U, \quad t + e^{\text{H}} = \sum_i
  n_i\langle\phi_i|\hat t|\phi_i\rangle + \frac{1}{2}\sum_{ij} n_i n_j
  \langle\phi_i\phi_j|r_{ij}^{-1}|\phi_i\phi_j\rangle,
\end{equation}
where $e^U$ is expressed through \emph{projection} onto correlated local
orbitals $|\bm R\mu\sigma)$ centered at site (or in the unit cell) $\bm R$ and
with orbital and spin quantum numbers $\mu$ and $\sigma$. The projection is
given by a local orbital occupation number $\tilde n_{\mu\sigma}$ which
depends on the variational quantities $\phi_i,\, n_i$:
\begin{equation}
  \label{eq:8}
  e^U = e^U(\tilde n_{\mu\sigma}[\phi_i,n_i]), \quad 
  \frac{1}{n_i}\,\frac{\delta}{\delta\phi^*_i} e_U = \sum_{\bm R\mu\sigma}
  \frac{\partial e^U}{\partial\tilde n_{\mu\sigma}}\frac{1}{n_i}\,
  \frac{\delta\tilde n_{\mu\sigma}}{\delta\phi^*_i}.
\end{equation}
The functional derivative on the r.h.s. of the last expression yields the
projection while the partial derivative defines the orbital and spin dependent
$U$-potential: $v^U_{\mu\sigma} = \partial e^U/\partial\tilde n_{\mu\sigma}$.
It is crucial for fitting the models in use into the general
Hohenberg-Kohn-Sham frame that \emph{the correlated orbitals themselves as
  well as the actual value of\/ $U$ are understood fixed and not variational}
although they may be context dependent. They kind of define a location,
relevant in a given context, in the variational functional space and a
functional contribution from that location.

\subsection{Correlated orbitals}

In the so called rotationally invariant LSDA+$U$ approach the correlated local
orbitals are assumed to be angular momentum eigenstates centered at $\bm R$,
say, with predefined orbital and spin momentum quantization axes (which both
need not be the same)
\begin{equation}
  \label{eq:9}
  |\bm R_i m_i \sigma_i), \quad m_i = -l_i,\ldots,l_i, \quad
  \sigma_i = \uparrow,\downarrow.
\end{equation}
Only one-site matrix elements, $\bm R_1 = \bm R_2 = \bm R_3 = \bm R_4$, are
considered:
\begin{equation}
  \label{eq:10}
  (m_1m_2|\tilde w|m_3m_4), \quad \tilde w \approx \tilde w(|\bm r - \bm r'|),
  \quad \sigma_1 = \sigma_3,\,\sigma_2 = \sigma_4
\end{equation}
The rotational invariance refers to the screened electron-electron
interaction, $\tilde w$, which is of course an approximation as regards the
screening. As a consequence, the $SO_3$ transformation properties of the
matrix elements are
\begin{multline}
  \label{eq:11}
  (m_1m_2|\tilde w|m_3m_4) =
\sum_{m'_1m'_2m'_3m'_4}
U^\dagger_{m_1m'_1}(\hat O)U^\dagger_{m_2m'_2}(\hat O)* \\
*(m'_1m'_2|\tilde w|m'_3m'_4)U_{m'_3m_3}(\hat O)U_{m'_4m_4}(\hat O),
\end{multline}
where $\hat O$ is any rotation of the $\bm r$-space and the $U$-matrices (not
to be confused with the Coulomb integral $U$) yield the relevant $SO_3$
representation:
\begin{equation}
  \label{eq:12}
  \begin{split}
    U^\dagger(\hat O)U(\hat O) =\; & 1 = U(\hat O)U^\dagger(\hat O), \\
    \int d\hat O\,U_{m_1m_2}(\hat O)U^\dagger_{m_3m_4}(\hat O) & =
    \frac{1}{2l+1}\,\delta_{m_1m_4}\delta_{m_2m_3}.
  \end{split}
\end{equation}
In the last orthogonality relation, $d\hat O$ is Haar's measure of the $SO_3$,
$\int d\hat O = 1$.

These fundamental representation properties allow for a very simple and
elegant derivation of the important sum rules for the matrix elements: Use
unitarity of $U$ and integrate over $d\hat O$ to obtain
\begin{equation}
  \label{eq:13}
  \begin{split}
    \lefteqn{\sum_{m_1} (m_1m_2|\tilde w|m_1m_4) =} & \\
    & =\sum_{m_1}\sum_{m'_1m'_2m'_3m'_4} U^\dagger_{m_1m'_1}(\hat O)
    U^\dagger_{m_2m'_2}(\hat O)
    \;(m'_1m'_2|\tilde w|m'_3m'_4)\cdot \\[-2ex]
    & \qquad\qquad\qquad\qquad\qquad\qquad\qquad\qquad\qquad
    \cdot U_{m'_3m_1}(\hat O)U_{m'_4m_4}(\hat O) = \\
    & = \sum_{m'_1m'_2m'_3m'_4} \delta_{m'_1m'_3}U^\dagger_{m_2m'_2}(\hat O)
    (m'_1m'_2|\tilde w|m'_3m'_4)U_{m'_4m_4}(\hat O) = \\
    & = \frac{1}{2l+1}\sum_{m'_1m'_2m'_4} (m'_1m'_2|\tilde w|m'_1m'_4)
    \delta_{m_2m_4}\delta_{m'_2m'_4} = \\
    & = \frac{\delta_{m_2m_4}}{2l+1} \sum_{m'_1m'_2}
    (m'_1m'_2|\tilde w|m'_1m'_2)\; = \delta_{m_2m_4}(2l+1)\; U.
  \end{split}
\end{equation}
The last equation is the definition of the Coulomb integral $U$. In the same
manner,
\begin{equation}
  \label{eq:14}
  \begin{split}
    \lefteqn{\sum_{m_1} (m_1m_2|\tilde w|m_3m_1) =} & \\
    & = \frac{\delta_{m_2m_3}}{2l+1} \sum_{m'_1m'_2}
    (m'_1m'_2|\tilde w|m'_2m'_1)\; = \delta_{m_2m_3}(U + 2lJ)
  \end{split}
\end{equation}
is obtained which additionally defines the exchange integral $J$. The first
result (\ref{eq:13}) is intuitively obvious: after summation over $m_1$ and
integration over $\bm r$ in the matrix element, no angular dependence with
respect to $\bm r'$ is left except the orthogonality $(m_2|m_4) =
\delta_{m_2m_4}$. The second result (\ref{eq:14}) is less obvious but
nevertheless true.

Expansion of the interaction function into spherical harmonics,
\begin{equation}
  \label{eq:15}
  \begin{split}
    \lefteqn{\tilde w\bigl(|\bm r_1 - \bm r_2|\bigr) =
    \tilde w\Bigl((r_1^2 + r_2^2 - 2r_1r_2\cos\theta)^{1/2}\Bigr) =} \\
    & = \sum_{l=0}^\infty \tilde w_l(r_1,r_2)\,P_l(\cos\theta) =
    \sum_{l=0}^\infty \tilde w_l(r_1,r_2)\frac{4\pi}{2l+1}
    \sum_{m=-l}^l Y_{lm}(\hat{\bm r}_1)Y_{lm}^*(\hat{\bm r}_2)
  \end{split}
\end{equation}
leads to Slater's analysis
\begin{equation}
  \label{eq:16}
  \begin{split}
    (m_1m_2|\tilde w|m_3m_4) &= 
    \sum_{l=0}^{2l_i}\tilde F_l\,a_l(m_1m_2m_3m_4), \\
    \tilde F_l &= \iint\limits_0^{\quad\infty} dr_1dr_2\,
    \bigl(r_1R_i(r_1)\bigr)^2\bigl(r_2R_i(r_2)\bigr)^2 \tilde w_l(r_1,r_2) \\
    & \approx \iint\limits_0^{\quad\infty} dr_1dr_2\,
    \bigl(r_1R_i(r_1)\bigr)^2\bigl(r_2R_i(r_2)\bigr)^2
    \frac{r_{<}^l}{r_{>}^{l+1}} \text{ for } l>0, \\
    a_l(m_1m_2m_3m_4) &= \frac{4\pi}{2l+1} \sum_{m=-l}^l
    (Y_{l_im_1}|Y_{lm}|Y_{l_im_3})(Y_{l_im_4}|Y_{lm}|Y_{l_im_2})^*.
  \end{split}
\end{equation}
Here, $l_i$ is the angular momentum of the considered shell, and the second
line for $\tilde F_l$ holds for the unscreened Coulomb interaction which for
$l>0$ is a reasonable approximation since intraatomic screening is effective
only for the $s$-component of the interaction.

Now, from $\sum_m Y_{lm}(\bm r)Y^*_{lm}(\bm r) = P_l(1)(2l+1)/4\pi$ and
\begin{equation*}
  \begin{split}
    \lefteqn{\sum_{m_1} a_l(m_1m_2m_1m_2) = } & \\
    & = \frac{4\pi}{2l+1} \Biggl[\sum_{m_1}
    (Y_{l_im_1}|Y_{l0}|Y_{l_im_1}) \Biggr]
    (Y_{l_im_2}|Y_{l0}|Y_{l_im_2})^* = \\
    & = \sqrt{4\pi}\frac{2l_i+1}{2l+1} \;\delta_{l0}\;
    (Y_{l_im_2}|Y_{l0}|Y_{l_im_2})^* = (2l_i+1)\delta_{l0}
  \end{split}
\end{equation*}
it follows immediately that
\begin{equation}
  \label{eq:17}
  U = \tilde F_0.
\end{equation}
Furthermore,
\begin{equation*}
  \begin{split}
    \sum_{m_1m_2} a_l(m_1m_2m_2m_1) & =
    \frac{4\pi}{2l+1} \sum_{\substack{ m_1m_2 \\ m }}
    (Y_{l_im_1}|Y_{lm}|Y_{l_im_2})(Y_{l_im_2}|Y^*_{lm}|Y_{l_im_1}) = \\
    & = \frac{4\pi}{2l+1} \iint d\Omega_1d\Omega_2
    \Bigl(\sum_{m_1} Y_{l_im_1}(\bm r_2)Y^*_{l_im_1}(\bm r_1)\Bigr)* \\
    & *\Bigl(\sum_{m_2} Y_{l_im_2}(\bm r_1)Y^*_{l_im_2}(\bm r_2)\Bigr)
    \Bigl(\sum_{m} Y_{lm}(\bm r_1)Y^*_{lm}(\bm r_2)\Bigr) = \\
    & = \frac{(2l_i+1)^2}{(4\pi)^2} \iint d\Omega_1d\Omega_2
    \bigl[P_{l_i}(\cos\theta_{12})\bigr]^2 P_l(\cos\theta_{12}) = \\
    & = \frac{(2l_i+1)^2}{4\pi} \int d\Omega
    \bigl[P_{l_i}(\cos\theta)\bigr]^2 P_l(\cos\theta) = \\
    & = (2l_i+1)^2 \begin{pmatrix} l_i & l & l_i \\ 0 & 0 & 0 \end{pmatrix}^2
  \end{split}
\end{equation*}
and hence
\begin{equation}
  \label{eq:18}
  \begin{split}
    \sum_{m_1m_2} (m_1m_2|\tilde w|m_2m_1) & = (2l_i+1)^2 \sum_{l=0}^{2l_i}
    \tilde F_l\;\begin{pmatrix} l_i & l & l_i \\ 0 & 0 & 0 \end{pmatrix}^2 =
    \\
    & = (2l_i+1)(U + 2l_i J).
  \end{split}
\end{equation}
Eqs. (\ref{eq:17}) and (\ref{eq:18}) relate the Coulomb and exchange integrals
$U$ and $J$ to Slater's (screened) integrals $\tilde F_l$.

Recall, however, that the whole analysis presupposes the isotropy of screening
which could be questioned at least in cases of strong directional covalency.

\subsection{The orbital occupation matrix}

The variants of the LSDA+$U$ model are all depending on the basis set of the
solver of the Kohn-Sham equations. There are a few subtleties in this game
which never have been discussed in the literature. Here, a non-orthogonal
local basis implementation \cite{Koe99} of the solver will be used, since a
local orbital representation is mandatory for considering strong correlations.
Non-orthogonality of the basis is rather the rule than the exception for high
precision solvers. For an LMTO solver see \cite{Lie95}, for an LAPW solver see
\cite{Shi99}.

Consider Kohn-Sham orbitals $|k\rangle = |\phi_k\rangle$ and orbital
occupation numbers $n_k$ as previously; they need not be eigenstates of spin.
Let $\{|l)\}$ be a possibly non-orthogonal basis for Kohn-Sham orbitals:
$|k\rangle = \sum_l|l)c_{lk},\; S_{ll'} = (l|l').$ (Systematically, brackets
are used for the Kohn-Sham orbitals and parentheses for the local basis
orbitals.) For an orthogonal projection onto those basis orbitals the
contragredient basis $|l\} = \sum_{l'} |l')(S^{-1})_{l'l},\; \{l|l') =
\delta_{ll'}$ is needed. With its help, the occupation matrix $\tilde n =
\tilde n[\phi_k,n_k]$ of correlated orbitals $|m\sigma)$ at site $\bm R$ in an
orthogonal form is introduced as
\begin{equation}
  \label{eq:19}
  \begin{split}
  \tilde n_{mm'\sigma} &=
    \sum_k\sum_{ll'} (S^{-1})_{(\bm R m\sigma),l}(l|k\rangle n_k
    \langle k|l')(S^{-1})_{l',(\bm R m'\sigma)} \;= \\
    &= \sum_k c_{(\bm R m\sigma),k} \; n_k \; c^*_{(\bm R m'\sigma),k}.
  \end{split}
\end{equation}
As usually it is assumed that the spin dependence can be made site diagonal by
choosing a suitable spin quantization axis. The orbital occupation matrix may
be diagonalized with respect to $m,m'$ at each lattice site $\bm R$ and for
each spin value $\sigma$ independently:
\begin{equation}
  \label{eq:20}
  \tilde n_{mm'\sigma} = \tilde U^{(\sigma)}_{m\mu_\sigma}\,
  \tilde n_{\mu\sigma}\, \tilde U^{(\sigma)*}_{m'\mu_\sigma}.
\end{equation}
Averages over a correlated shell of angular momentum $l$,
\begin{equation}
  \label{eq:21}
  \tilde n_\sigma = \frac{1}{2l+1} \sum_\mu \tilde n_{\mu\sigma}, \quad
  \tilde n = \frac{1}{2} \bigl(\tilde n_\uparrow + \tilde n_\downarrow\bigr),
\end{equation}
are used later on.

The projector in (\ref{eq:8}) is now
\begin{equation}
  \label{eq:22}
  \frac{1}{n_k}\frac{\delta\tilde n_{\mu\sigma}}{\delta\langle k|} =
  \sum_{ll'} |l')(S^{-1})_{l',(\bm R \mu\sigma)}
  (S^{-1})_{(\bm R \mu\sigma),l}(l|k\rangle
\end{equation}
Naturally, in applications the correlated orbitals are assumed to form a
subset of the basis orbitals, although this is not mandatory. In the FPLO
scheme, the basis is adjusted in the course of iterations for solving the
non-linear Kohn-Sham equations. This does not mean that the basis itself is
treated as variational. Rather the relevant sector of the variational space is
tracked along the way of search for the Kohn-Sham minimum. Likewise, the
relevant location of correlation, that is the correlated orbitals as part of
the basis, is tracked along.

\subsection{The Orbital Polarization LSDA+$U$ Functional}
 
This functional was introduced under the name `around the mean field' (AMF) in
Ref.~\cite{Czy94}. It is zero if the orbitals of an atomic shell are equally
occupied, hence it depends on orbital polarization. It is given by
\begin{equation*}
  l(n_{ss'}(\bm r),\ldots) = l_{\text{LSDA}},
\end{equation*}
\begin{equation}
  \label{eq:23}
  \begin{split}
    e^{U,\text{AMF}} & = \frac{1}{2}\sum_{\bm R\sigma\mu\mu'} \biggl\{
    (\mu_\sigma\mu'_{-\sigma}|\tilde w|\mu_\sigma\mu'_{-\sigma})
    (\tilde n_{\mu\sigma} - \tilde n_\sigma)
    (\tilde n_{\mu'-\sigma} - \tilde n_{-\sigma})
    \; + \\[-2ex]
    & + \bigl[(\mu_\sigma\mu'_\sigma|\tilde w|\mu_\sigma\mu'_\sigma) -
    (\mu_\sigma\mu'_\sigma|\tilde w|\mu'_\sigma\mu_\sigma)\bigr]
    (\tilde n_{\mu\sigma} - \tilde n_\sigma)
    (\tilde n_{\mu'\sigma} - \tilde n_\sigma)
    \biggr\} \\
    & = \frac{1}{2}\sum_{\bm R\sigma\mu\mu'} \biggl\{
    (\mu_\sigma\mu'_{-\sigma}|\tilde w|\mu_\sigma\mu'_{-\sigma})\;
    \tilde n_{\mu\sigma}\tilde n_{\mu'-\sigma} + \\[-4ex]
    & \qquad\qquad\qquad
    + \bigl[(\mu_\sigma\mu'_\sigma|\tilde w|\mu_\sigma\mu'_\sigma) -
    (\mu_\sigma\mu'_\sigma|\tilde w|\mu'_\sigma\mu_\sigma)\bigr]\;
    \tilde n_{\mu\sigma}\tilde n_{\mu\sigma}\biggr\} \\[-2ex]
    & \qquad - \frac{1}{2}\sum_{\bm R\sigma}
    \biggl\{U\bigl(N - \tilde n_\sigma\bigr) -
    J\Bigl(N_\sigma - \tilde n_\sigma\Bigr)\biggr\}N_\sigma, \\
    & N_\sigma = \sum_\mu \tilde n_{\mu\sigma} = (2l+1)\tilde n_\sigma.
  \end{split}
\end{equation}
Here, $N$ is the number of electrons occupying a whole correlated $l$-shell,
$N_\sigma$ is that for one spin sort. There is no danger of confusing it with
the total electron number in Eqs. (\ref{eq:1}--\ref{eq:5}), the latter does
not appear any more in the sequel. In the second equality use of the sum rules
for the matrix elements was made.

The corresponding $U$-potential is, again most easily with use of the sum
rules,
\begin{multline}
  \label{eq:24}
  \frac{\partial e^{U,\text{AMF}}}{\partial\tilde n_{\mu\sigma}} = \sum_{\mu'}
  \biggl\{(\mu_\sigma\mu'_{-\sigma}|\tilde w|\mu_\sigma\mu'_{-\sigma})
  (\tilde n_{\mu'-\sigma} - \tilde n_{-\sigma}) + \\
  + \bigl[(\mu_\sigma\mu'_\sigma|\tilde w|\mu_\sigma\mu'_\sigma) -
  (\mu_\sigma\mu'_\sigma|\tilde w|\mu'_\sigma\mu_\sigma)\bigr]
  (\tilde n_{\mu'\sigma} - \tilde n_\sigma) \biggr\}
\end{multline}
One weak point of this version is that it yields no contribution at all in
case of orbital independent occupation numbers $\tilde n_{\mu'\sigma} = \tilde
n_\sigma$. This is for instance the case of a half-filled completely spin
polarized shell (e.g. $4f$-shell of Gd). In the Gd case this is at least not
too bad, as the LSDA gives nearly the right spin polarization energy of Gd,
although there is a problem with the right magnetic ground state (obtained
antiferromagnetic in LSDA).

\subsection{The `Atomic Limit' LSDA+$U$ Functional}

With the Gd case in mind and aiming at a better description of the
photoelectron spectra, Czy\.zyk and Sawatzky introduced another functional in
\cite{Czy94} which they labeled `atomic limit' (AL). At least regarding its
relation to photoemission it should rather be considered a model for the
quasiparticle self-energy $\Sigma$ instead of being related to $e^U$.
Nevertheless it was given in an $e^U$-form as (again $l(n_{ss'}(\bm r),\ldots)
= l_{\text{LSDA}}$)
\begin{equation}
  \label{eq:25}
  \begin{split}
    & e^{U,\text{AL}} = \frac{1}{2}\sum_{\bm R\sigma\mu\mu'} \Biggl\{
    (\mu_\sigma\mu'_{-\sigma}|\tilde w|\mu_\sigma\mu'_{-\sigma})\;
    \tilde n_{\mu\sigma}\tilde n_{\mu'-\sigma} + \\[-4ex]
    & \qquad\qquad\qquad\qquad
    + \bigl[(\mu_\sigma\mu'_\sigma|\tilde w|\mu_\sigma\mu'_\sigma) -
    (\mu_\sigma\mu'_\sigma|\tilde w|\mu'_\sigma\mu_\sigma)\bigr]\;
    \tilde n_{\mu\sigma}\tilde n_{\mu\sigma}\Biggr\} \\[-2ex]
    & \qquad - \frac{1}{2}\sum_{\bm R} \biggl\{UN\bigl(N - 1\bigr) -
    J\sum_\sigma N_\sigma\Bigl(N_\sigma - 1\Bigr)\biggr\}, = \\
    & = e^{U,\text{AMF}} +
    \frac{1}{2}\sum_{\bm R\sigma} (U-J)(1-\tilde n_\sigma)N_\sigma.
  \end{split}
\end{equation}
The corresponding $U$-potential is
\begin{equation}
  \label{eq:26}
  \frac{\partial e^{U,\text{AL}}}{\partial\tilde n_{\mu\sigma}} =
  \frac{\partial e^{U,\text{AMF}}}{\partial\tilde n_{\mu\sigma}} -
  (U-J)\Bigl(\tilde n_\sigma - \frac{1}{2}\Bigr).
\end{equation}
One characteristic feature of this $U$-potential is that in case of an
isolated shell it moves the occupied states downward by $(U-J)/2$ and the
unoccupied states upward by $(U-J)/2$ independent of the shell occupation. By
way of contrast, the center of the AMF spin subshell potential split moves up
with increasing subshell occupation (so that the shift of the occupied levels
is zero in the case of a filled spin subshell and likewise the shift of the
unoccupied levels of an empty spin subshell; this way yielding no shift at all
in Gd.) On the other hand, the AL $4f$-level splitting of Gd is approximately
doubled compared to LSDA which is a rather good result in the sense of a
self-energy correction.

\subsection{The Kohn-Sham Hamiltonian matrix element and the total energy}

Here the formulas are presented, which actually are implemented. For the sake
of simplicity, the correlated orbitals are identified with selected local
basis orbitals as discussed at the beginning of subsection 2.2. The
diagonalization of the occupation matrix, although greatly simplifying the
analytical derivations, does not have advantages when coded, since anyway the
diagonalizing transformation is site- and spin-dependent. Therefore, the full
occupation matrix $\tilde{n}_{mm^\prime\sigma}$ is kept. (In the
following, the site index of all quantities is dropped. If multiple sites with
correlated states are needed, the formulas apply to all sites separately.)
The matrix is obtained from the KS states.
\begin{equation}
  \tilde{n}_{mm^\prime\sigma}=\sum_{n \bm k} c^{n \bm k}_{m\sigma}
  n_{n\bm k\sigma}c^{n\bm k*}_{m^\prime\sigma}
\end{equation}
Integration is over the irreducible part of the Brillouin zone and a
symmetrization projector is applied afterwards to get the result for the full
zone.

The AMF $U$-potential matrix as given in the text after Eq. (\ref{eq:8})
becomes
\begin{equation}
v^{\mathrm{AMF}}_{m^\prime m\sigma} 
 = \sum_{\sigma^\prime}
\sum_{\mu\mu^\prime}
(\tilde{n}_{\mu\mu^\prime\sigma^\prime}
-\tilde{n}_{\sigma^\prime} \delta_{\mu\mu^\prime})
\left [ 
 (m^\prime\mu^\prime|\tilde{w}|m\mu)
-\delta_{\sigma\sigma^\prime}
 (m^\prime\mu^\prime|\tilde{w}|\mu m)
\right ]
\end{equation}
with the property ${\mathrm{Tr}} \ v^{\mathrm{AMF}}_{mm^\prime\sigma} \equiv
0$ for each $\sigma$ separately. There is no such property of the AL potential
since here the up and down shifts are independent of the shell occupation. The
interaction matrix elements are taken to be spin independent. They are
calculated from the Slater parameters according to Eq. (\ref{eq:16}). (Recall
that the $\tilde{F}_i$ are external parameters, not variational.)

The projection part of Eq. (\ref{eq:8}) gives only Kronecker deltas. Thus the
matrix elements of the KS equation are modified by
$v^{\mathrm{AMF}}_{mm^\prime\sigma}$ for every block of correlated orbitals.
From the eigenvalues the band structure energy $E^B$ is calculated as usual.
The kinetic energy is obtained from it by subtracting double counting
corrections. Besides the LSDA corrections the band energy contains a term
\begin{equation}
  \Delta T^{\mathrm{AMF}} =\sum_{m m^\prime \sigma} 
  v^{\mathrm{AMF}}_{m^\prime m\sigma} \tilde{n}_{m m^\prime\sigma}.
\end{equation}
Since the $U$-potential matrix is traceless, a constant diagonal term may be
added to the occupation matrix to obtain
\begin{equation}
  \Delta T^{\mathrm{AMF}} =\sum_{m m^\prime \sigma} 
  v^{\mathrm{AMF}}_{m^\prime m\sigma} (\tilde{n}_{m m^\prime\sigma}-
  \tilde{n}_{\sigma} \delta_{m m^\prime})=2E^{U,{\mathrm{AMF}}}.
\end{equation}
Finally, the LSDA+$U$, kinetic and total energy are
\begin{eqnarray*}
  E^{U,{\mathrm{AMF}}} &= & \frac{1}{2} \Delta T^{\mathrm{AMF}}\\
  T             &= & T^{\mathrm{LSDA}}-2 E^{U,{\mathrm{AMF}}}\\
  E             &= & E^{\mathrm{LSDA}}-E^{U,{\mathrm{AMF}}}.
\end{eqnarray*}
The $U^{\text{AL}}$-potential matrix becomes
\begin{equation}
v^{\mathrm{AL}}_{m^\prime m\sigma} 
 = \sum_{\sigma^\prime}
\sum_{\mu\mu^\prime}
\tilde{n}_{\mu\mu^\prime\sigma^\prime}
\left [ 
 (m^\prime\mu^\prime|\tilde{w}|m\mu)
-\delta_{\sigma\sigma^\prime}
 (m^\prime\mu^\prime|\tilde{w}|\mu m)
\right ]
-v^{\mathrm{dc}}_{\sigma} \delta_{m m^\prime}
\end{equation}
with 
\begin{equation}
  v^{\mathrm dc}_{\sigma} = U(N-\frac{1}{2}) -J(N_{\sigma} -\frac{1}{2}).
\end{equation}
Again, every block belonging to a correlated orbital in the Hamilton
matrix is modified by adding $v^{\mathrm{AL}}_{mm^\prime\sigma}$.

Using a nutritive zero it can be shown that
\begin{equation}
  \Delta T^{\mathrm{AL}} \equiv \sum_{m m^\prime \sigma} 
  v^{\mathrm{AL}}_{m^\prime m\sigma}
  \tilde{n}_{m m^\prime\sigma} = 2E^{U,{\mathrm{AL}}} -\frac{U-J}{2}N
\end{equation}
Thus, the final result is
\begin{eqnarray*}
  E^{U,{\mathrm{AL}}} &= & \frac{1}{2} \Delta T^{\mathrm{AL}} +
  \frac{1}{2}\frac{U-J}{2}N\\
  T             &= & T^{\mathrm{LSDA}}-2 E^{U,{\mathrm{AL}}}+\frac{U-J}{2}N\\
  E             &= & E^{\mathrm{LSDA}}-E^{U,{\mathrm{AL}}}+\frac{U-J}{2}N.
\end{eqnarray*}

\section{Applications to Cuprates}

The structure of the 2D cuprates considered here may be described by the
formula B(CuO$_2$)$_n$Ca$_{n-1}$, where B denotes the block layer which
separates stacks of $n$ CuO$_2$-planes with Ca planes sandwiched in between.

The LSDA+$U$ approach was applied to three compounds (CaCuO$_2$ ($n=\infty$),
Sr$_2$CuO$_2$Cl$_2$ ($n=1$) and Bi$_2$Sr$_2$CaCu$_2$O$_8$ ($n=2$)) with the
focus on the orbital analysis of relevant bands. While the bandstructure gap
between occupied and unoccupied bands is a ground state property (jump of the
chemical potential as function of the particle number at zero temperature),
the whole bandstructure refers to the excitation spectrum, and the KS
bandstructure need not compare to photoemission, say. Nevertheless, although
there is no deeper reason that the best LSDA+$U$ potential for the KS
equations should be close to the electron self-energy, there is some hope that
like for weakly correlated systems the LSDA+$U$ bandstructure could again
provide also an approximation to the quasi-particle spectrum. One should,
however, be aware that in principle the best KS $U$-value need not be the same
as the best $U$-value in Hubbard-type model Hamiltonians. The latter value
should for instance be used in dynamical mean-field theory which is an
approach to the electron Green's function (self-energy). A satisfactory link
of the $U$-value to the KS variational quantities, the KS orbitals and KS
orbital occupation numbers, is still missing. For LSDA+$U$ as a Hohenberg-Kohn
model the most relevant results are ground state properties as structural
parameters (lattice constants, Wyckoff parameters), magnetic structure and
magnetic polarization energy and band gap. The lattice parameters for the
cuprates are obtained in the usual $\pm$2 p.c. agreement with experiment. They
will not be considered in the following.

The FPLO version used is 3.00-5 \cite{fplo}.  Here, the program settings are
summarized, which are unique to all calculations.  The LSDA version is that of
Perdew and Zunger \cite{Per81}.  The Cu $3d$ orbitals are taken to be the
correlated orbitals.  For the sake of comparison, the same Slater parameters
were used for all calculations: $U=8.16$ eV ($0.3$ Hartree), $J=1$ eV ($F_2=9$
eV, $F_4=5$ eV). The orbitals were optimized in the non-magnetic structures
and the resulting compression radii were used also for the antiferromagnetic
(AFM) LSDA-$U$ calculations. Unless explicitly else stated the ``around mean
field'' (AMF) functional is used.

\subsection{CaCuO$_2$}
 
\begin{figure}
  \begin{center}
    \includegraphics[scale=0.4]{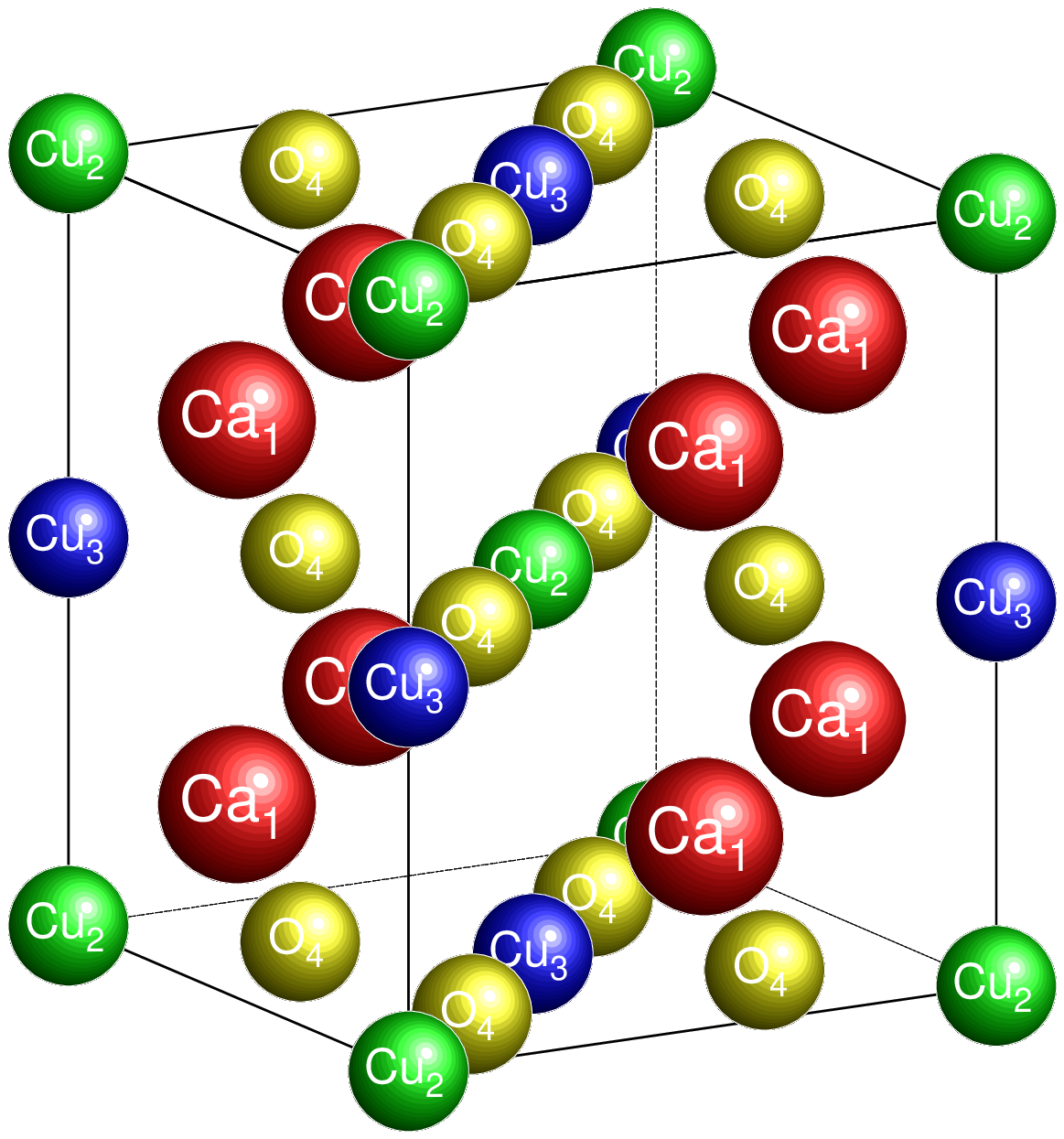}\hspace{1cm}
    \includegraphics[scale=0.4]{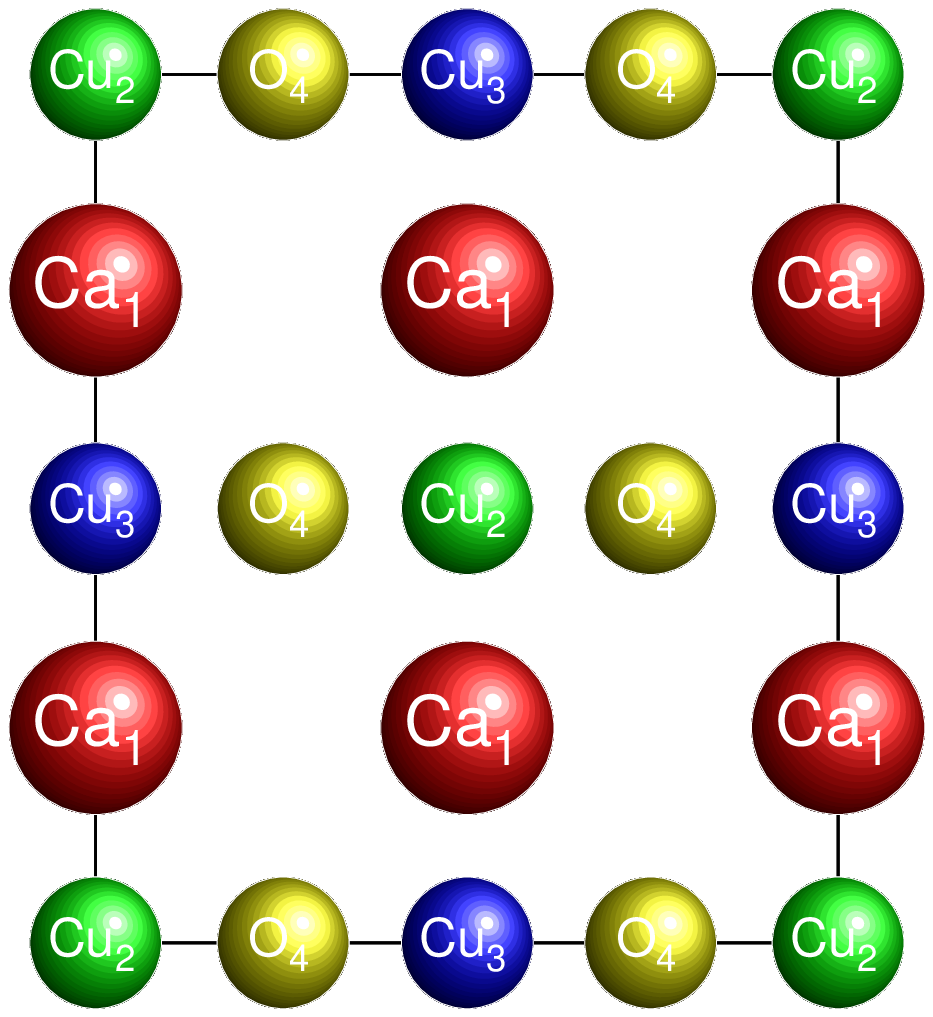}
    \caption{Unit cell of antiferromagnetic CaCuO$_2$; Cu$_2$ spin up, Cu$_3$
      spin down.}
    \label{fig2}
  \end{center}
\end{figure}

The `infinite layer' cuprate CaCuO$_2$ (no block layer B) does not exist in
nature, but there is an isostructural compound Ca$_{0.85}$Sr$_{0.15}$CuO$_2$.
It is considered here first because the bands of this infinite cuprate stack
are not perturbed by hybridization with block layer states and hence in this
sense are pure.  The experimental lattice parameters of
Ca$_{0.85}$Sr$_{0.15}$CuO$_2$ \cite{Vak89} are taken for the fictitious
CaCuO$_2$. The antiferromagnetic unit cell is shown in Fig.~\ref{fig2}. The
space group is I4/mmm (139). The distance along Cu-O-Cu is $d_a=3.86$
\AA\ and the distance in $z$-direction between adjacent CuO$_2$ layers is
$d_c=3.20$ \AA. (For all compounds considered here $d_a$ is the lattice
constant of the non-magnetic cuprate plane, while $d_c$ is the distance in
$z$-direction between adjacent cuprate multilayers [(CuO$_2$)$_n$Ca$_{n-1}$].
For CaCuO$_2$ it is the CuO$_2$ plane distance.  It also gives the periodicity
of the multilayers in $z$-direction, ignoring a centering shift perpendicular
to the $z$-direction.)

The AFM lattice constants are $a_0=b_0=\sqrt{2}d_a$, $c_0=2 d_c$. The atom
positions are Ca $(0,\frac{1}{2},\frac{1}{4})$, Cu $(0,0,0)$, Cu
$(0,0,\frac{1}{2})$ and O $(\frac{1}{4},\frac{1}{4},0)$.
Table~\ref{tab:BasCaCuO2} gives the FPLO basis set. The number of Fourier
components was $1024$ per atom and the k-mesh subdivision was $(6,6,6)$.
Nonrelativistic calculations were performed.

The symmetry points of the band structures presented below refer to the
Brillouin zone of the antiferromagnetic cuprate plane (a square of edge length
$2\pi/(\sqrt{2}d_a)$). For better comparison to the literature we relate it to
the non-magnetic Brillouin zone which is a rotated by 45$^o$ square of edge
length $2\pi/d_a$. In units of $\pi/d_a$ we have $\Gamma,(Z)=(0,0,\zeta)$,
$X,(R)=(\frac{1}{2},\frac{1}{2},\zeta)$ and $M,(A)=(1,0,\zeta)$. The first
labels refer to $\zeta=0$ and the labels in parentheses refer to $\zeta = 1
\;\hat=\; \pi/d_c$.  (This is the $Z$ point of a simple tetragonal cell of
lattice constant $c=d_c$.)

\begin{figure}
  \begin{center}
    \includegraphics[scale=0.25]{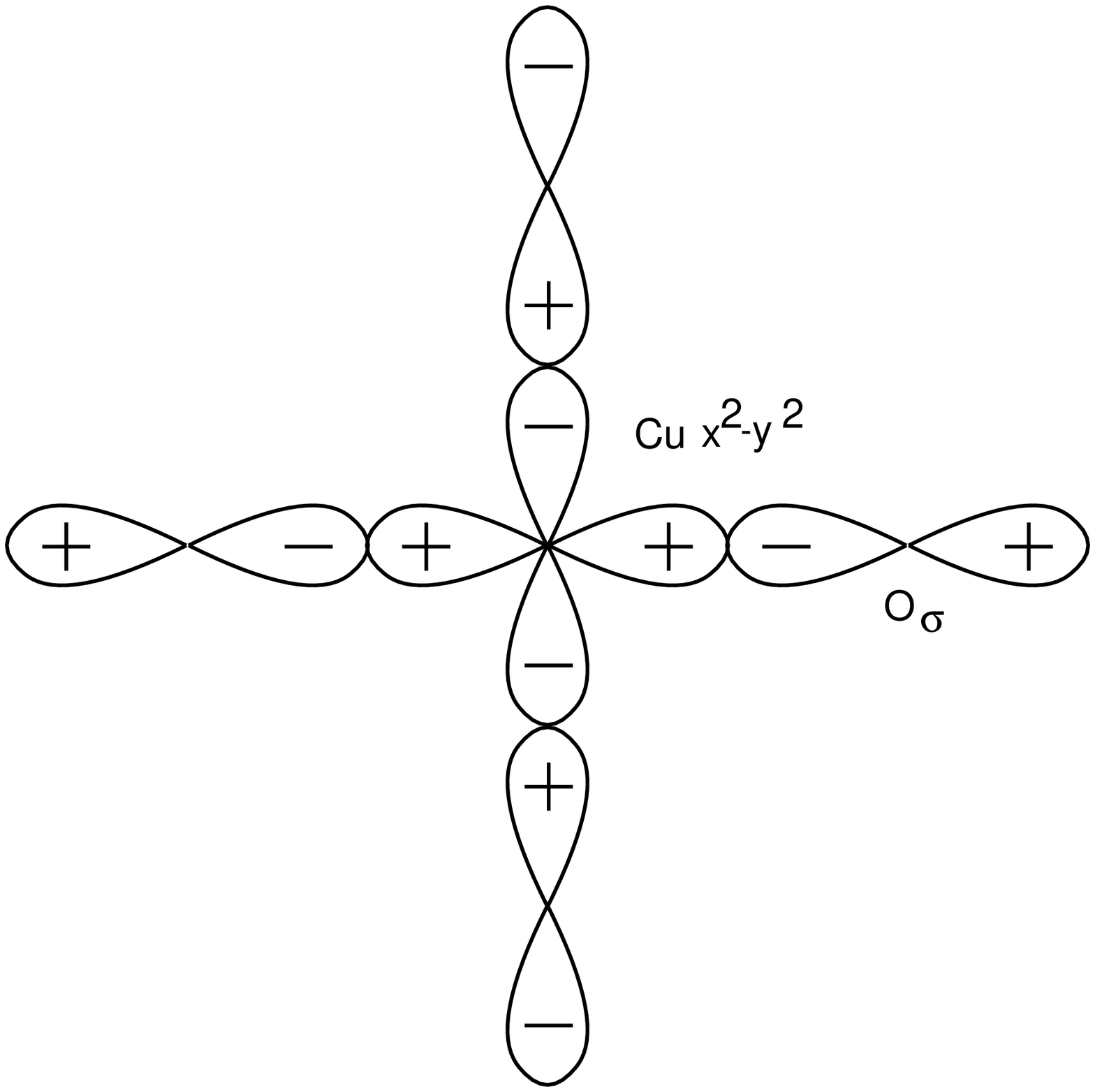}\hspace{1cm}
    \includegraphics[scale=0.25]{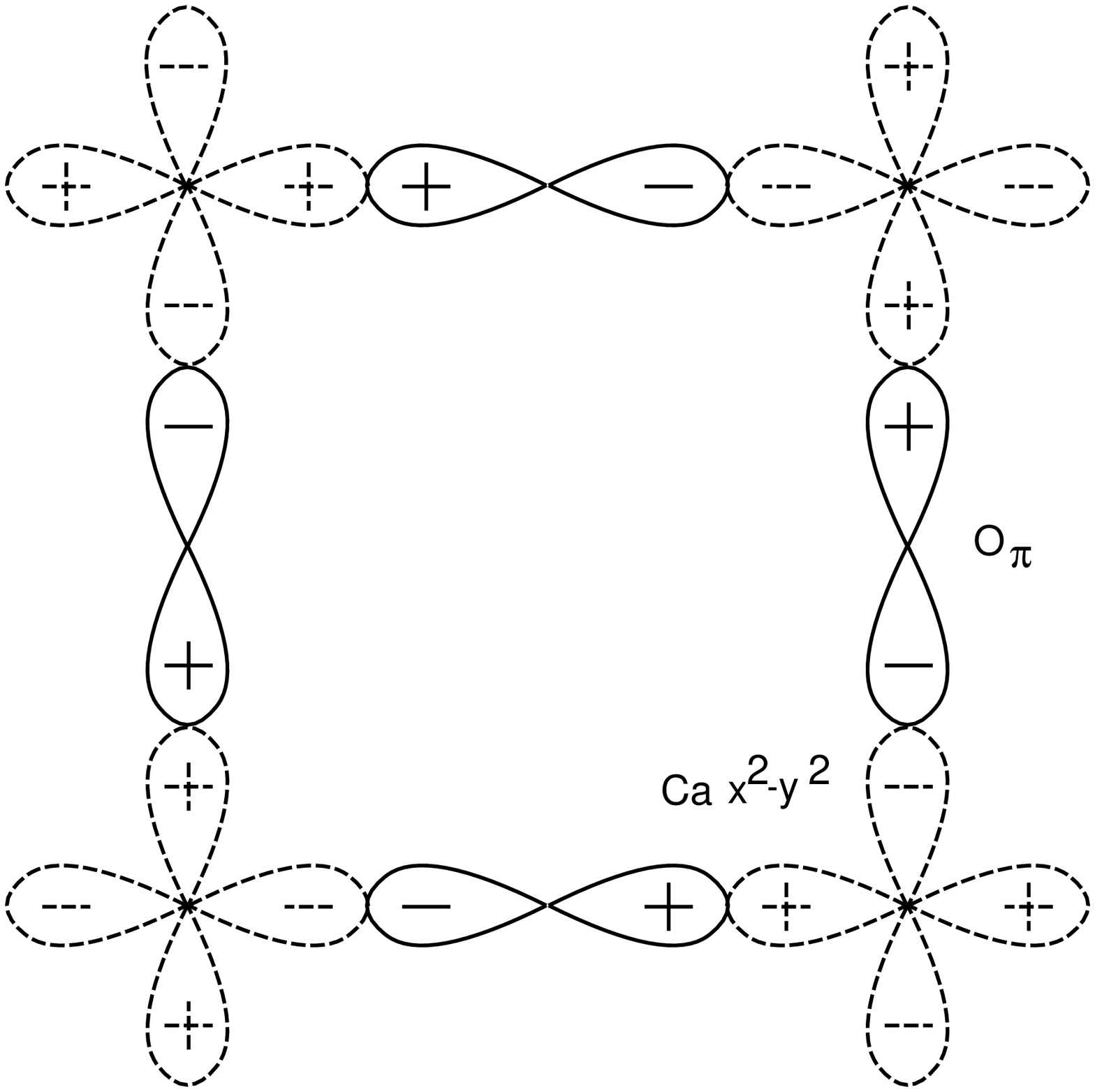}
    \caption[HOMO]{Left: Antibonding ($\bm k = (\pi,\pi,0)$) $dp\sigma$-orbital
      commonly assumed as the HOMO that forms the Zhang-Rice singlet together
      with the nominal Cu-$d$ hole \cite{Eme87,Zha88}.\\
      Right: O-O antibonding ($\bm k = 0$) in-plane $p\pi$-orbital, lifted up
      by crystal field and weakly hybridized with Ca-$d$ orbitals in adjacent
      layers: the true HOMO of the LSDA+$U$ model \cite{Pot97,Hay99}.}
    \label{fig3}
  \end{center}
\end{figure}

Calculations within LSDA and LSDA+$U$ were performed. As usual for the
cuprates, the LSDA gives a metallic ground state. The KS bandstructure is
shown in Fig.~\ref{fig4lsda}. The spaghetti below the Fermi level (here and in
all following figures put equal to zero) consists of hybridized Cu-$3d$ and
O-$2p$ states, with the bonding combinations at the bottom and the antibonding
bands formed of the orbitals of Fig.~\ref{fig3} crossing the Fermi
level. Non-bonding combinations are in between. The unoccupied bands above the
Fermi level start with Cu-$4s$ and Cu-$4p$ character and then enter a bunch of
Ca-$3d$ bands above 5~eV. 

The same bands weighted (by linewidth) with the square of the coefficient of
selected basis orbitals in the KS state are shown in Fig.~\ref{fig5lsda}. From
the upper two panels one can read off a (Cu-O)$_\sigma$ covalency split of
more than 5~eV while the third panel shows a (O-O) covalency split of about
3~eV. One further observes that the O$_\sigma$ orbitals and the O$_z$ orbitals
hybridize also with the Cu $4s$ and $4p$ orbitals while the in-plane O$_\pi$
orbitals hybridize additionally with the Ca $3d$ orbitals. Figs.~\ref{fig4lsda}
and \ref{fig5lsda} are presented here for comparison with the LSDA+$U$ results
shown below.

Experimentally, (Ca$_{0.85}$Sr$_{0.15}$)CuO$_2$ is an AFM insulator with a
bandgap of more than 1 eV and a N\'eel temperature $T_N \approx 540$ K. The
$U$-functional cures this deficiency and one finds an AFM solution with a spin
polarization energy $\Delta E=27.9 \mathrm{mHa}$ per formula unit below the
Pauli-paramagnetic (PM) state. The site projected copper $3d$ moment is $0.71
\mu_B$ and the total copper spin moment is $0.69 \mu_B$ (reduced by negative
$3s3p$ moments).

The already discussed two relevant molecular orbitals (MO) of Fig.~\ref{fig3}
are the candidates for the highest occupied molecular orbital (HOMO) in the
correlated electronic structure of cuprates, that is, those MOs which are
relevant for the valence band edge. The LSDA+$U$ KS bands of CaCuO$_2$ with
the AMF and AL functionals are shown in Fig.~\ref{fig4}. The main difference
between both functionals is found in the unoccupied bands. As to be expected
for a more than half-filled shell, the upper Hubbard band of
(Cu-$3d_{x^2-y^2}$ and O-$2p_{\sigma}$ character, see below) lies higher in
the AMF case compared to AL. In AL this band is the lowest unoccupied band. In
the occupied part near the Fermi level the differences are small, while far
below larger differences are found due to the different position of the lower
Hubbard band and thus to different hybridization. Since this paper focuses on
the occupied bands near the Fermi level, the differences are not very
relevant, and all further results are presented for the AMF functional.

On Fig.~\ref{fig5} the orbital weights to the AMF bands in analogy to
Fig.~\ref{fig5lsda} is shown. First, it is clearly seen that the upper and
lower Hubbard band is formed by Cu-$3d_{x^2-y^2}$ and O-$2p_\sigma$ orbitals.
At the valence band edge (points X and R shown, but without noticeable
dispersion on the whole line X-R in $z$-direction of the $\bm k$-space)
O-$2p_\sigma$ and O-$2p_\pi$ contribute equally strongly while the
Cu-$3d_{x^2-y^2}$ orbital contribution is largely suppressed compared to the
LSDA result. This suppression of the Cu contribution to valence holes in
cuprates is confirmed by experiment \cite{Fin94}. The O-$2p_\pi$ contribution
on the other hand is strongly enhanced compared to LSDA, a new result which is
missed in most model Hamiltonian treatments in the literature where the
O-$2p_\pi$ degree of freedom is excluded from the Hamiltonian in most cases.
Other orbitals do not contribute. Hence, on the line X-R the highest occupied
band is mainly a hybrid of O-$2p_{\sigma}$ and O-$2p_{\pi}$. These band states
are assumed to form the Zhang-Rice singlet with the nominal $3d$-hole on the
Cu site (upper Hubbard state) \cite{Zha88}. The bandwidth of the highest
valence band is about 1~eV due to hybridization with a flat band 1~eV below
the Fermi level. On the line X-M (R-A) the O-$2p_{\sigma}$ contribution fades
away towards M (A) due to this hybridization. In LSDA the O-$2p_{\pi}$
orbitals do practically not contribute to this band which led to the neglect
of that orbital in model Hamiltonians. This failure of the LSDA is mainly due
to the fact that in LSDA the O-$2p_{\pi}$ bands are deeper in energy compared
to the Cu-$3d_{x^2-y^2}$--O-$2p_\sigma$ bands.

The absolutely highest occupied band in the LSDA+$U$ result is a pure
O-$2p_{\pi}$ band (at $\Gamma$) which therefore is in energetic competition
with the Zhang-Rice state when adding additional holes. The O-$2p_{\pi}$ bands
show a considerable dispersion in $z$-direction which comes from a weak
hybridization with the unoccupied Ca-$3d_{x^2-y^2}$ orbitals. The Cu $4s$
orbitals do not contribute at the valence band edge. They are mixed into the
unoccupied states and are also slightly mixed into O-$2p_{\sigma}$ states
about 1.5~eV below the Fermi level on the line M-A.

\subsection{Sr$_2$CuO$_2$Cl$_2$}

The single layer compound Sr$_2$CuO$_2$Cl$_2$ is probably the most
two-dimensional of all cuprates. The block layer consists of 2 SrCl layers
which separate single CuO$_2$ planes. Between these SrCl layers crystals are
easily cleaved, whence most photoemission data on undoped planar cuprates are
recorded from this material. Two adjacent cuprate planes are shifted
horizontally by a shift vector $(\frac{1}{2},\frac{1}{2},0)$ relative to each
other, which produces a body centered non-magnetic unit cell. The large
distance of cuprate planes from each other prevents valence state coupling in
$z$-direction. The AFM cell has base centered orthorhombic symmetry and is
shown in Fig.~\ref{fig6}.

The standard cell choice for the orthorhombic cell, which has also to be used
in the FPLO code results in c-base centering. (The stacking direction is the
$b$-direction and the cuprate plane is the $c$,$a$ plane.) The space group is
Cmmm (65).  The cell parameters are $d_a=3.973$ \AA{} and $2 d_c=15.618$ \AA\ 
\cite{Mil90}. The lattice constants of the AFM cell are $a_0=c_0=\sqrt{2}d_a$,
$b_0=2 d_c$.  The atom positions are Cu $( 0 ,\frac{1}{2},\frac{1}{2})$, Cu $(
0 ,0 ,0)$, O $( -\frac{1}{4}, 0, -\frac{1}{4})$, Cl $( 0,
-0.317,\frac{1}{2})$, Cl $( 0, 0.183, 0)$, Sr $( 0, -0.107, \frac{1}{2})$ and
Sr $( 0, 0.393, 0)$.  Table~\ref{tab:BasSrCuOCl} gives the basis set.  The
number of Fourier components was $1000$ per atom and the k-mesh subdivision
was $(12,12,12)$. Scalar relativistic calculations are performed. For the sake
of comparison the stacking direction is further on denoted $z$ and the cuprate
plane as the $x,y$-plane. The symmetry points are chosen corresponding to the
scheme described in the previous subsection.

The site projected copper $3d$ moment of the LSDA+$U$ result is $0.758 \mu_B$
and the total copper spin moment is $0.748 \mu_B$.  In Fig.~\ref{fig7} the
orbital weights for the relevant bands are shown.  The situation is rather
similar to that of CaCuO$_2$, so only the differences are pointed out.

The ``Zhang-Rice'' band is slightly higher and touches the Fermi level at X
and R. The intersecting band is correspondingly lower at the line M-A, as a
consequence the width of the upper valence band is about 1.5~eV and the fading
of the O-$2p_{\sigma}$ character on the lines X-M and R-A is less pronounced,
a reduction of about 50 p.c. remains. The $z$-dispersion of the O-$2p_{\pi}$
band on the line $\Gamma$-Z has gone as there are no Ca-$3d$ states present
for hybridization. There is a marked 2D character of the compound.

The quasiparticle low energy dispersion measured by ARPES (single hole
excitation) is shown in Fig.~\ref{fig9}. A detailed discussion is found in
\cite{Toh00} and \cite{Dam03}.  The left part of the experimental spectra
(from (0,0) to ($\frac{\pi}{2}$,$\frac{\pi}{2}$)) compares nicely with the
O-$2p_\sigma$ dominated LSDA+$U$ band on the line $\Gamma-X$ (second
panel of Fig.~\ref{fig7}) and the right part (from
($\frac{\pi}{2}$,$\frac{\pi}{2}$) to (0,$\pi$)) with the same LSDA+$U$ band on
the line $X-M$.  Even the reported fading ARPES intensity when going from $X$
towards $M$ (see also \cite{Ron02}) agrees with the fading O-$2p_\sigma$
projection of that band. Nevertheless, the experimental band width is smaller
by a factor of about two to three and the situation on the line $\Gamma-M$ is
less clear although the comparison of only the O-$2p_\sigma$ projected bands
(second panel of Fig.~\ref{fig7}, cf. the discussion of models above) to model
results contained in Fig.~\ref{fig9} is not so bad. After all, LSDA+$U$
accounts for electron correlations still rather grossly.

\subsection{Bi$_2$Sr$_2$CaCu$_2$O$_8$}

The bilayer compound Bi$_2$Sr$_2$CaCu$_2$O$_8$ has a double layer of cuprate
planes separated by block layers. The block layer consists of two BiSrO$_2$
layers between which crystals cleave equally easily as in the previous case
and also favor the material for photoemission. Adjacent double layers are
again shifted horizontally by a shift vector $(\frac{1}{2},\frac{1}{2},0)$
relative to each other, which produces a body centered tetragonal non-magnetic
cell. The hypothetic AFM cell again is assumed base centered orthorhombic.

The cell parameters are again given in the c-base centered setting, resulting
in the $b$-axis being the stacking direction.  The space group is Cmmm (65).
The cell parameters are $d_a=3.817$ \AA{} and $2 d_c=30.6$ \AA{}.  The lattice
constants of the AFM cell are $a_0=c_0=\sqrt{2}d_a$, $b_0=2 d_c$.  The atom
positions are Ca $( 0, \frac{1}{4} \frac{1}{4} )$, Sr $( 0, 0.3597
,\frac{1}{4})$, Sr $( 0, 0.1403 ,\frac{1}{4})$, Bi $( 0, -0.4478
,\frac{1}{4})$, Bi $( 0, -0.0522 ,\frac{1}{4})$, Cu $( 0, -0.3040
,\frac{1}{4})$, Cu $( 0, -0.1960 ,\frac{1}{4})$, O$_1$ $( -\frac{1}{4},
0.1960, 0)$, O$_2$ $( 0, -0.1250 ,\frac{1}{4})$, O$_2$ $( 0, -0.3750
,\frac{1}{4})$, O$_3$ $( 0, 0.0450 ,\frac{1}{4})$ and O$_3$ $( 0, 0.4550
,\frac{1}{4})$ . The two Sr, Bi, O$_2$, O$_3$ and Cu atoms are equivalent in
the non-magnetic cell. The O$_1$ atom is that of the cuprate plane. The two Cu
spins are antiferromagnetically ordered in the AFM cell.
Table~\ref{tab:BasBiSrCaCuO} gives the basis set. The number of Fourier
components was $500$ per atom and the k-mesh subdivision was $(8,8,8)$. Again,
scalar relativistic calculations were performed and the symmetry points in
$\bm k$-space are chosen corresponding to the scheme described in subsection
3.2.

For this material both the LSDA and LSDA+$U$ yield a metallic ground state
with Cu-O bands and Bi bands crossing the Fermi level. While the LSDA results
in a non-magnetic solution, the LSDA+$U$ calculation yields a stable AFM
state.  The site projected copper $3d$ moment is $0.696 \mu_B$ and the total
copper spin moment is $0.684 \mu_B$.

Fig.~\ref{figBi} shows both the LSDA and LSDA+$U$ bands of
Bi$_2$Sr$_2$CaCu$_2$O$_8$ close to the Fermi level. For a better orientation
in the band character the Cu-$3d_{x^2-y^2}$ projected bands are also shown.
Note that the LSDA+$U$ results on the right panels have twice as many bands as
the LSDA results on the left panels due to the AFM order of the former ground
state. The bands crossing the Fermi level and not seen in the lower panels are
Bi bands (more precisely BiO bands hybridized with orbitals of the block layer
oxygen).

The splitting of the Cu-$3d_{x^2-y^2}$ projected LSDA bands crossing the Fermi
level on the line X-M (R-A) is the much discussed bilayer splitting between
bonding and antibonding combinations of the CuO states in both CuO$_2$ layers
of the bilayer \cite{Dam03}. The coupling of those states is mainly due to a
small hybridization with Cu-$4s$ states, and this part of the coupling has
$k_x^2-k_y^2$ symmetry. Hence, the splitting is maximum ($\sim$ 0.25 eV) at
point M (A) and nearly zero at point X (R) although a very small splitting
remains there due to a small direct coupling. It is readily seen by a simple
symmetry argument that in the AFM state this splitting must be zero on the
whole line X-M (R-A), if spin-orbit coupling is neglected. Accordingly all
bands on the right panels of Fig.~\ref{figBi} are twofold degenerate on these
lines and do not show a bilayer splitting there. This only develops away from
these lines, for instance in the X-$\Gamma$ and M-$\Gamma$ directions.
However, the Cu-$3d_{x^2-y^2}$ projected LSDA+$U$ bands on the right lower
panel show around M and A another splitting of approximately the same
magnitude which is due to the crossing of another oxygen band. This point has
never been considered in the literature to date. 

Note also that the non-magnetic structure of the left panels has a larger
Brillouin zone so that the rising band from $\Gamma$ to X continues to rise
from X=($\frac{1}{2},\frac{1}{2}$,0) to the Brillouin zone corner (1,1,0),
whereas the AFM structure of the right panels has a charge transfer gap
(dominated by $U$ and hence much larger than the exchange splitting) at X,
which now lies on the Brillouin zone boundary, and the conduction band has its
maximum at X. On the other hand, the band pair close to the Fermi level on the
line $\Gamma-Z$ on the upper right panel is on the line (1,1,$\zeta$) in the
non-magnetic state and not shown on the upper left panel.  To illustrate the
charge transfer gap, the LSDA+$U$ band structure projected on
Cu-$3d_{x^2-y^2}$ is shown once more in Fig.~\ref{figBiCuBig} for a larger
energy window.

Experimentally, the oxygen in the block layers of Bi$_2$Sr$_2$CaCuO$_8$ is
volatile and its stoichiometry is governed by thermodynamics. Moreover, the
geometry of the block layer is distorted in a disordered way compared to the
ideal structure used in the calculation. In the recent, highest resolution 
photoemission spectra \cite{Dam03}, Fermi surface pockets around point M which
should be present due to the Bi bands in both the LSDA and LSDA+$U$ results
are not seen. Hence, one could assume that they are pushed away from the Fermi
level (together with the oxygen bands on $\Gamma$-Z) by a distortion
potential. Correspondingly, the Fermi level would be lowered to ensure the
electron count. This would be a big problem for the LSDA band structure where
not only the antibonding bilayer split band would have an electron Fermi
surface closed around $\Gamma$ but also the bonding one, both in contradiction
to what is seen in photoemission. The LSDA+$U$ band structure (lower right
panel of Fig.~\ref{figBi}) on the contrary would be in rather good agreement
with photoemission if one wipes out the down folded bands of the AFM order
which is not found in experiment. (The material cannot be reduced down to
undoped CuO$_2^{2-}$ planes.) 

There would be another stark difference to what is discussed in the literature
with respect to the bilayer splitting: If the band splitting around M and
observed in photoemission would be a bilayer splitting, one should expect it to
be strongly reduced when reducing the doping level to an underdoped
superconductor which is regarded to develop strong AFM correlations (also seen
in neutron scattering). It should be reduced to zero where AFM order sets in.
The splitting of the LSDA+$U$ bands of different origin on the contrary is to
be expected largely independent of antiferromagnetic order and hence on
doping.

\subsection{Implications on Magnetic Interactions}

Magnetic couplings may grossly be obtained from total energy differences of
LSDA+$U$ results for ferromagnetic and antiferromagnetic order; in more detail
they may be obtained from calculated energies of spin spiral states
\cite{Yar02}. 

For a more detailed understanding of their physics, a tight-binding model of
the kind of Emery's model should be extracted from the LSDA+$U$ results, which
then may be down-mapped to a kind of a $t-J$ model. For a hypothetic
ferromagnetic order (assumed for the sake of simplicity) such a tight-binding
model was derived in Ref.~\cite{Hay99}.

One main conclusion from the present orbital analysis is that in a large part
of the Brillouin zone there is a strong hybridization of O-2$p_{\sigma}$ with
O-2$p_{\pi}$ orbitals in bands hybridized with the Cu-3$d_{x^2-y^2}$ orbital.
Furthermore, the O-2$p_{\pi}$ bands are in energetic competition with the
bands forming the Zhang-Rice state, if additional holes are doped. Hence, the
O-2$p_{\pi}$ orbitals must be included in the Emery model in order to
correctly describe the $t$-terms which determine the magnetic coupling.

In CaCuO$_2$ there is a sizable dispersion of the in-plane O-2$p_{\pi}$ bands
in $z$-direction mediated by some hybridization with Ca-3$d_{x^2-y^2}$
orbitals.  This is in accordance with the experimental finding that CaCuO$_2$
has the highest N\'eel temperature, $T_N = 540$ K, of all layered cuprates
indicating 3D magnetism \cite{Vak89,Poz97}. For a survey of the magnetic
properties of cuprates see Ref.~\cite{Joh97}.

By contrast, in Sr$_2$CuO$_2$Cl$_2$ there is no dispersion of the
corresponding bands in $z$-direction due to the Sr$_2$Cl$_2$ buffer layers.
There is only dipole-dipole coupling of the planes, compatible with the
experimental findings, $T_N = 256$ K \cite{Gre94}.

\section{Conclusions}

The LSDA+$U$ approach is shown to fit perfectly in the frame of DFT by
Hohenberg and Kohn for the electronic ground state, provided the theory can be
closed by linking the $U$-value to the variational quantities, the KS orbitals
and orbital occupation numbers (which explicit link is yet to be rendered). On
the other hand, the LSDA+$U$ potential is also widely understood as an
approximation to the electron self-energy (first step towards an
LSDA+dynamical mean-field theory). There is hope that in this way like in the
situation of weakly correlated systems it provides again a tool to obtain a
rather accurate ground state and a reasonable approximation to the
quasi-particle spectrum (band structure) in a single run from only one set of
equations. To pursue this goal, in the first part of the present paper an
attempt was made to present the structure of the LSDA+$U$ theory as clearly as
possible.

Application to several typical planar cuprate structures and comparison to
experimental data of the isolating electronic state and the magnetic state as
well as of quasi-particle spectra probed by photoemission seems to support
this expectation. On the other hand it revealed a number of new aspects in the
physics of the electronic structure of cuprates. Notably the sufficient
completeness of most model Hamiltonians in use must be questioned, at least in
connection with the dimensionality and details of magnetic couplings, and the
so-called bilayer splitting of the band structure of Bi$_2$Sr$_2$CaCu$_2$O$_8$
should be reanalyzed.

\section{Acknowledgments}

We thank H. Rosner for helpful discussions. Financial support by the German
Israel Foundation under Contract No. I-614-13.14/99 and by the Deutsche
Forschungsgemeinschaft, SFB 463, is gratefully acknowledged.

\newpage

\begin{table}[htbp]
  \centering
  \begin{tabular}[t]{lll}
    \hline
    Atom & core & valence \\
    \hline
    \hline
    Ca & $1s$ $2s$ $2p$ & $3s$(-1) $3p$(-1) $4s$(1.1240) $4p$(1.0927) $3d$(-1)\\
    Cu & $1s$ $2s$ $2p$ & $3s$(-1) $3p$(-1) $4s$(1.3148) $4p$(1.2648) $3d$(1.3500)\\
    O  & $1s$           & $2s$(1.2869) $2p$(1.2822) $3d$(-1)\\
    \hline
  \end{tabular}
  \caption{Basis set for CaCuO$_2$. (Compression parameter for valence
    orbitals in parentheses.)}
  \label{tab:BasCaCuO2}
\end{table}
\begin{table}[htbp]
  \centering
  \begin{tabular}[t]{lll}
    \hline
    Atom & core & valence \\
    \hline
    \hline
    Cu & $1s$ $2s$ $2p$ & $3s$(-1) $3p$(-1) $4s$(1.2300) $4p$(1.1199) $3d$(1.3148)\\
    O  & $1s$           & $2s$(1.2580) $2p$(1.2377) $3d$(1.1162)\\
    Cl & $1s2s2p$       & $3s$(1.1518) $3p$(1.1338) $3d$(1.0852)\\
    Sr & $1s$ $2s$ $2p$ $3s$ $3p$ $3d$ & $4s$(-1) $4p$(-1) $5s$(1.1360) $5p$(1.0816) $4d$(1.1503) \\
    \hline
  \end{tabular}
  \caption{Basis set for Sr$_2$CuO$_2$Cl$_2$. (Compression parameter for valence
    orbitals in parentheses.)}
  \label{tab:BasSrCuOCl}
\end{table}

\begin{table}[htbp]
  \centering
  \begin{tabular}[t]{lll}
    \hline
    Atom & core & valence \\
    \hline
    \hline

    Ca & $1s$ $2s$ $2p$ & $3s$(-1) $3p$(-1) $4s$(1.1298) $4p$(1.1210) $3d$(1.2286) \\
    Sr & $1s$ $2s$ $2p$ $3s$ $3p$ $3d$ & $4s$(-1) $4p$(-1) $5s$(1.1239) $5p$(1.0717) $4d$(1.1435) \\
    Bi & $1s$ $2s$ $2p$ $3s$ $3p$ $3d$ $4s$ $4p$ $4d$ $4f$ & $5s$(-1) $5p$(-1) $6s$(1.3570) $6p$(1.2922) $5d$(1.4752) \\
    Cu & $1s$ $2s$ $2p$ & $3s$(-1) $3p$(-1) $4s$(1.2229) $4p$(1.1549) $3d$(1.3060) \\
    O$_1$  & $1s$ & $2s$(1.2464) $2p$(1.2177) $3d$(-1) \\
    O$_2$ & $1s$ &  $2s$(1.1766) $2p$(1.1895) $3d$(-1) \\
    O$_3$ & $1s$ & $2s$(1.0683) $2p$(1.0951) $3d$(-1) \\
    \hline
  \end{tabular}
  \caption{Basis set for Bi$_2$Sr$_2$CaCu$_2$O$_8$. (Compression parameter for
    valence orbitals in parentheses.)}
  \label{tab:BasBiSrCaCuO}
\end{table}
\begin{figure}
  \begin{center}
    \includegraphics[scale=.27,angle=-90]{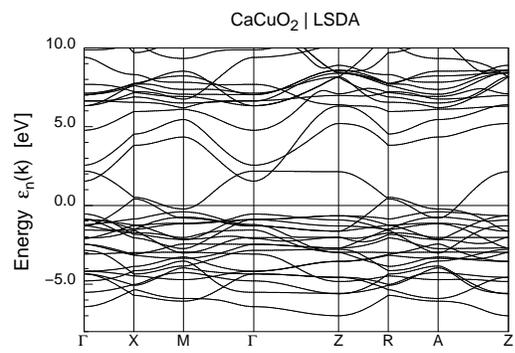}
    \caption{LSDA band structure of CaCuO$_2$.}
    \label{fig4lsda}
  \end{center}
\end{figure}
\begin{figure}
  \begin{center}\vspace*{-1cm}

    \includegraphics[scale=.27,angle=-90]{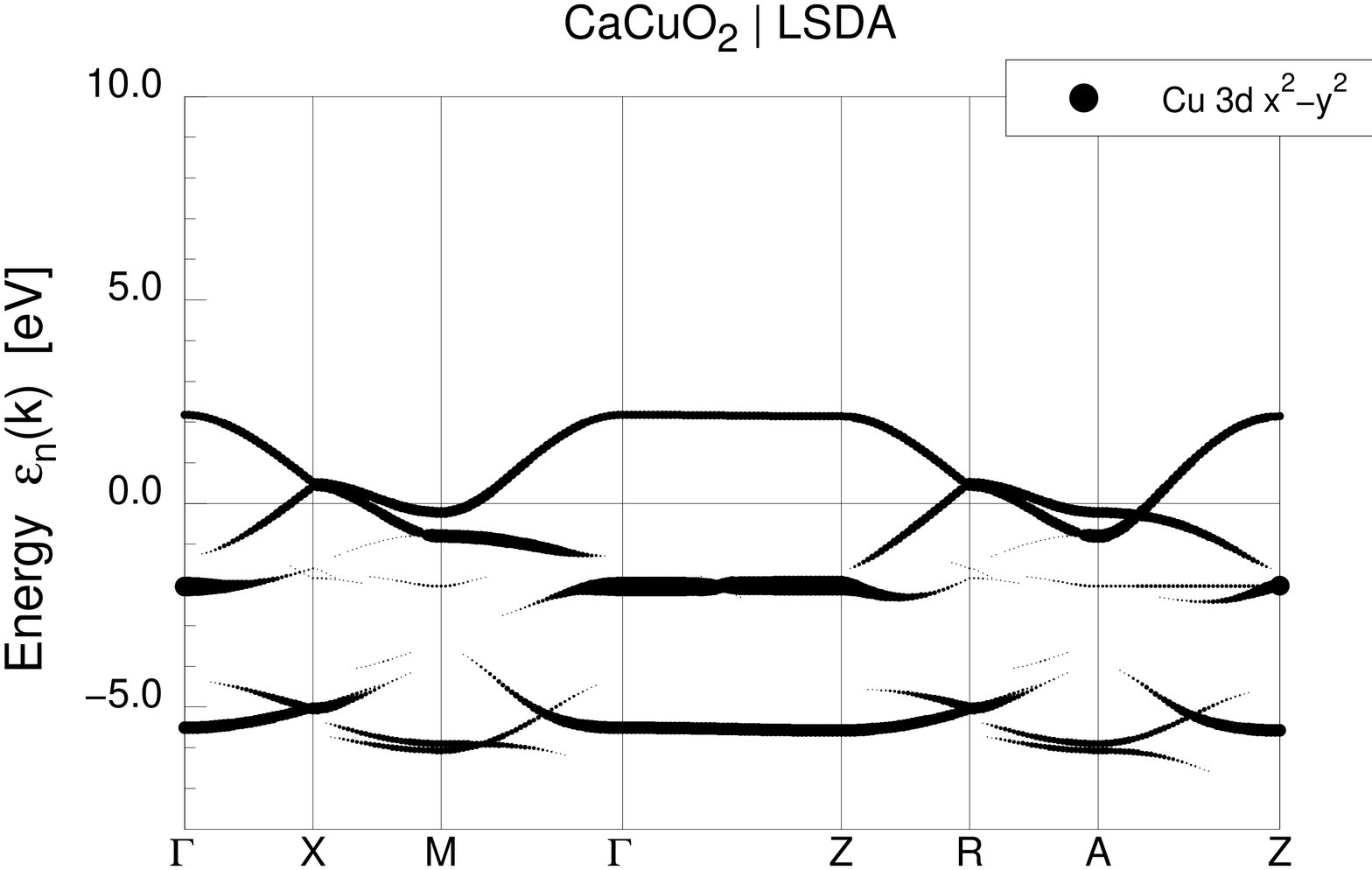}
    \vspace{-1cm}

    \includegraphics[scale=.27,angle=-90]{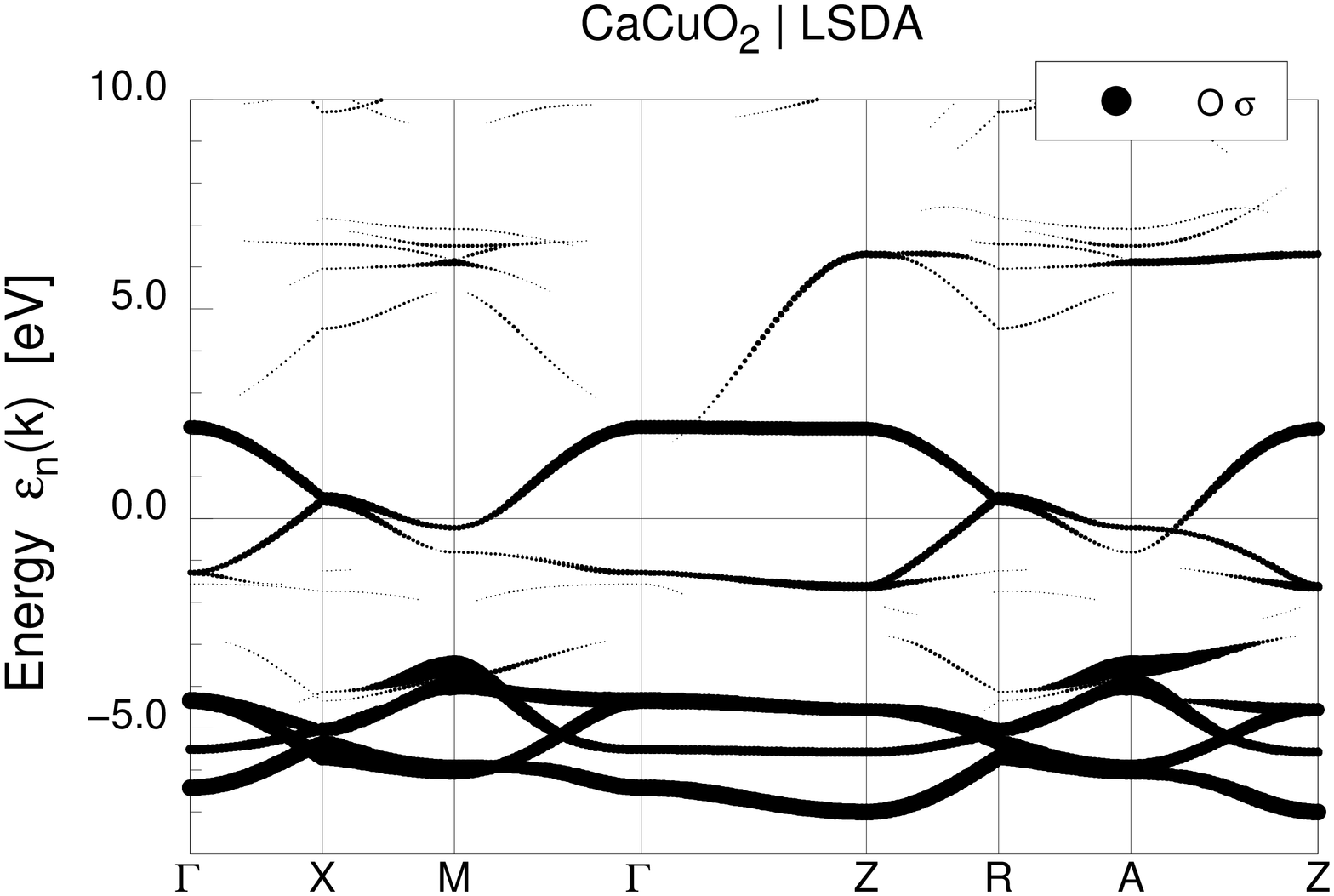}
    \vspace{-1cm}

    \includegraphics[scale=.27,angle=-90]{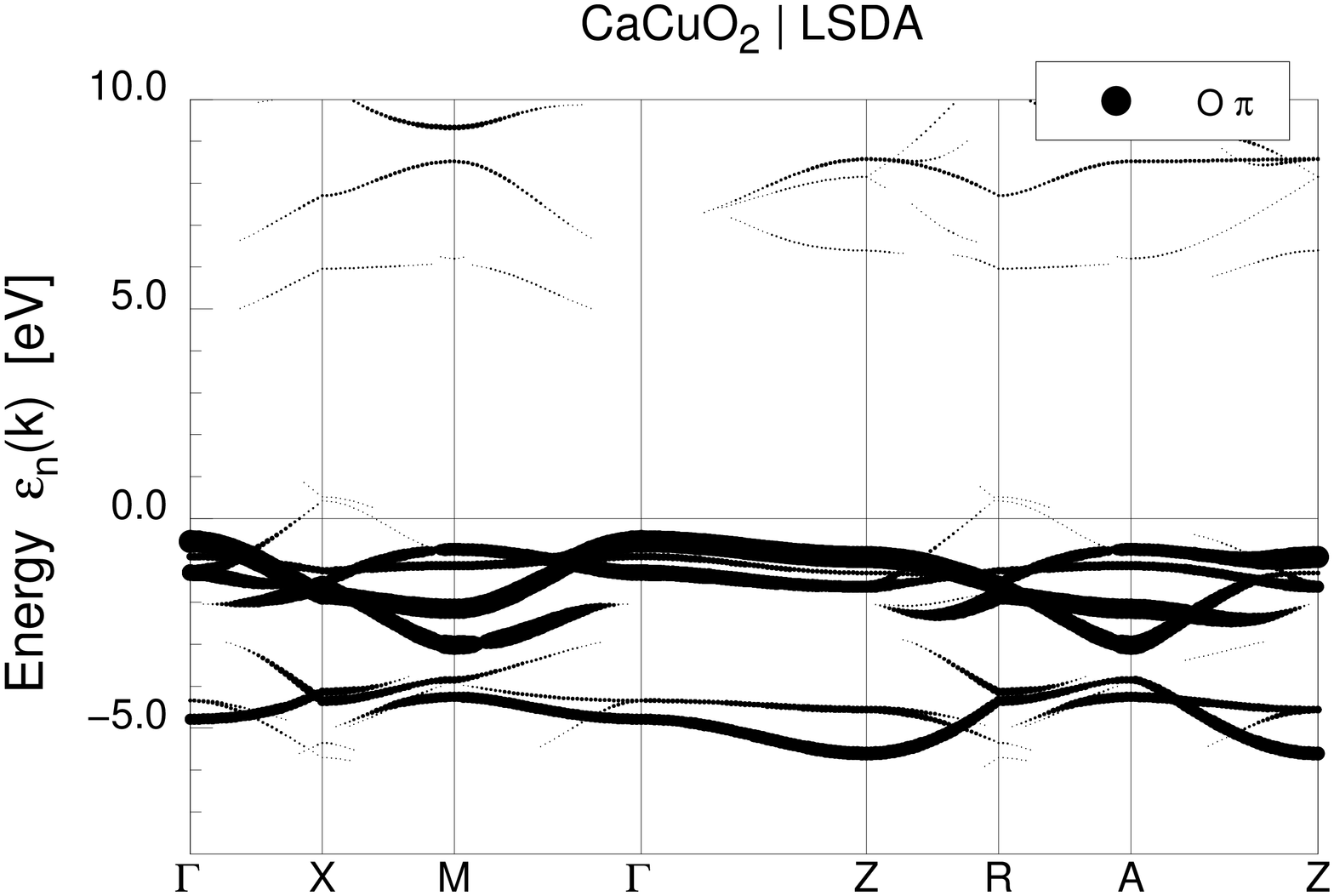}
    \vspace{-1cm}

    \includegraphics[scale=.27,angle=-90]{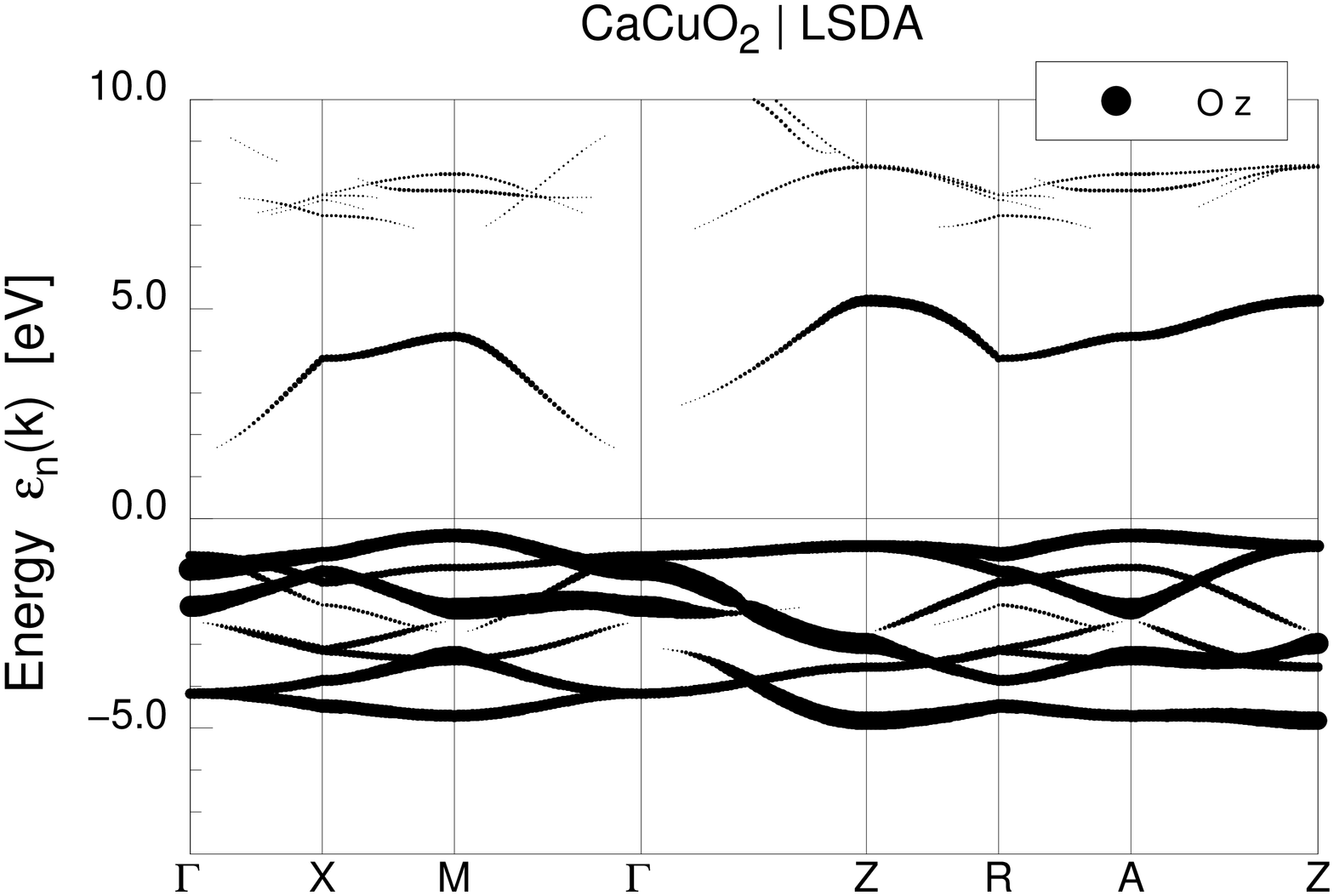}

    \caption{LSDA result for CaCuO$_2$. From top: orbital weight of the
      Cu-$3d_{x^2-y^2}$ orbital, the O-$2p_{\sigma}$ orbitals, the
      O-$2p_{\pi}$ and the O-$2p_z$ orbitals.} 
    \label{fig5lsda}
  \end{center}
\end{figure}
\begin{figure}
  \begin{center}\vspace*{-1cm}

    \includegraphics[scale=.27,angle=-90]{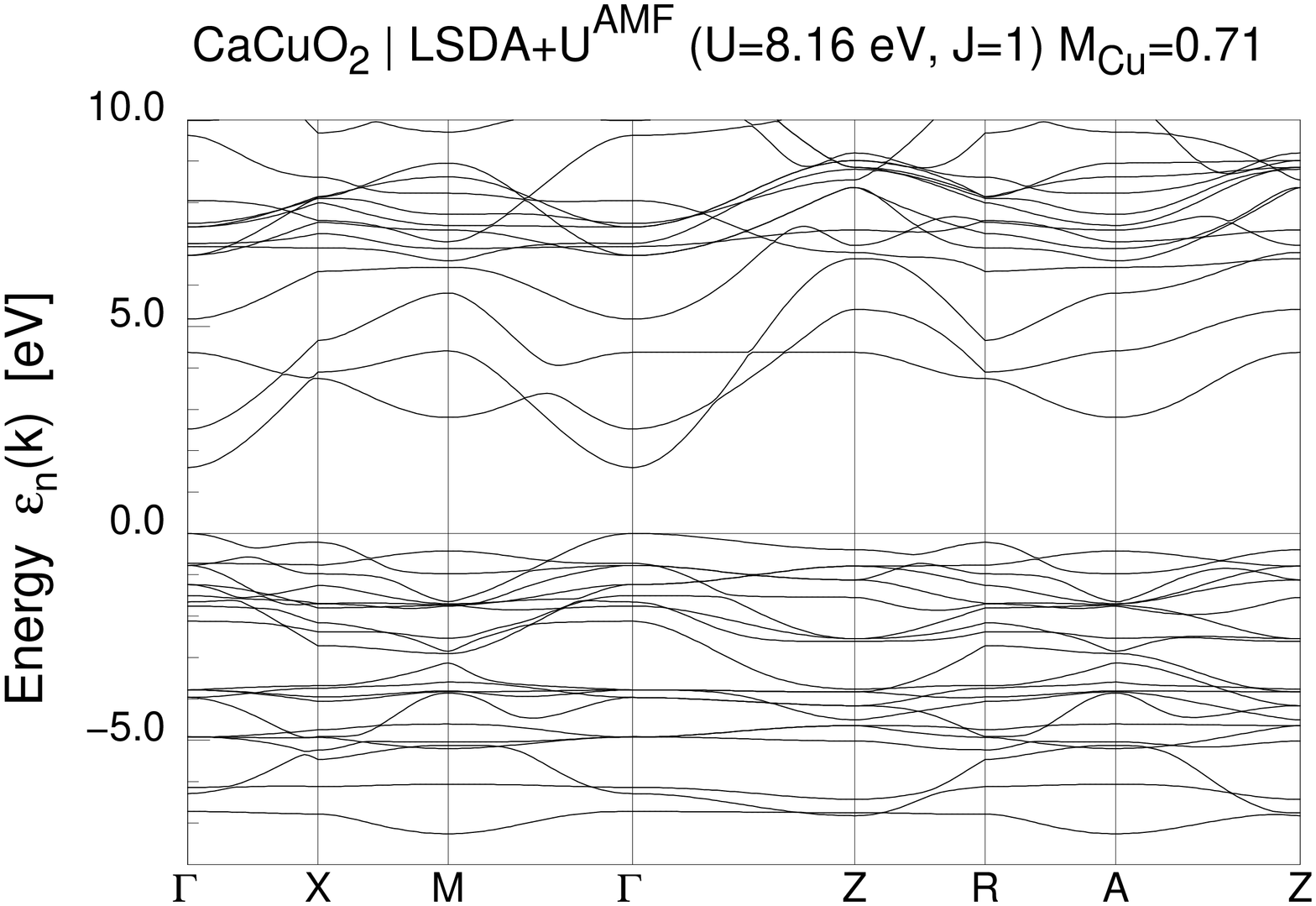}
    \vspace{-1cm}

    \includegraphics[scale=.27,angle=-90]{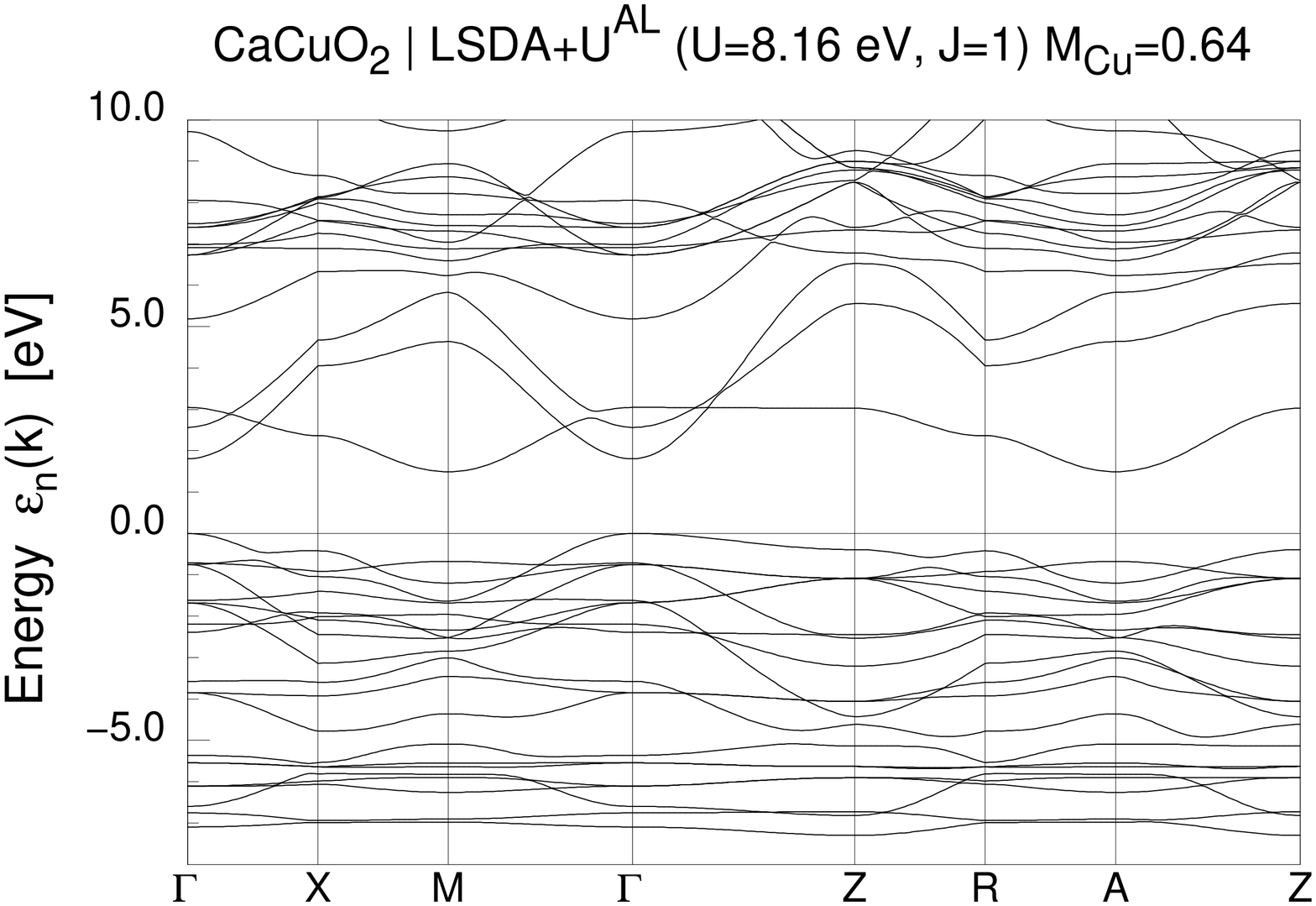}

    \caption{LSDA+$U$ band structure of CaCuO$_2$. Top: AMF, bottom: AL.}
    \label{fig4}
  \end{center}
\end{figure}
\begin{figure}
  \begin{center}\vspace*{-1cm}

    \includegraphics[scale=.27,angle=-90]{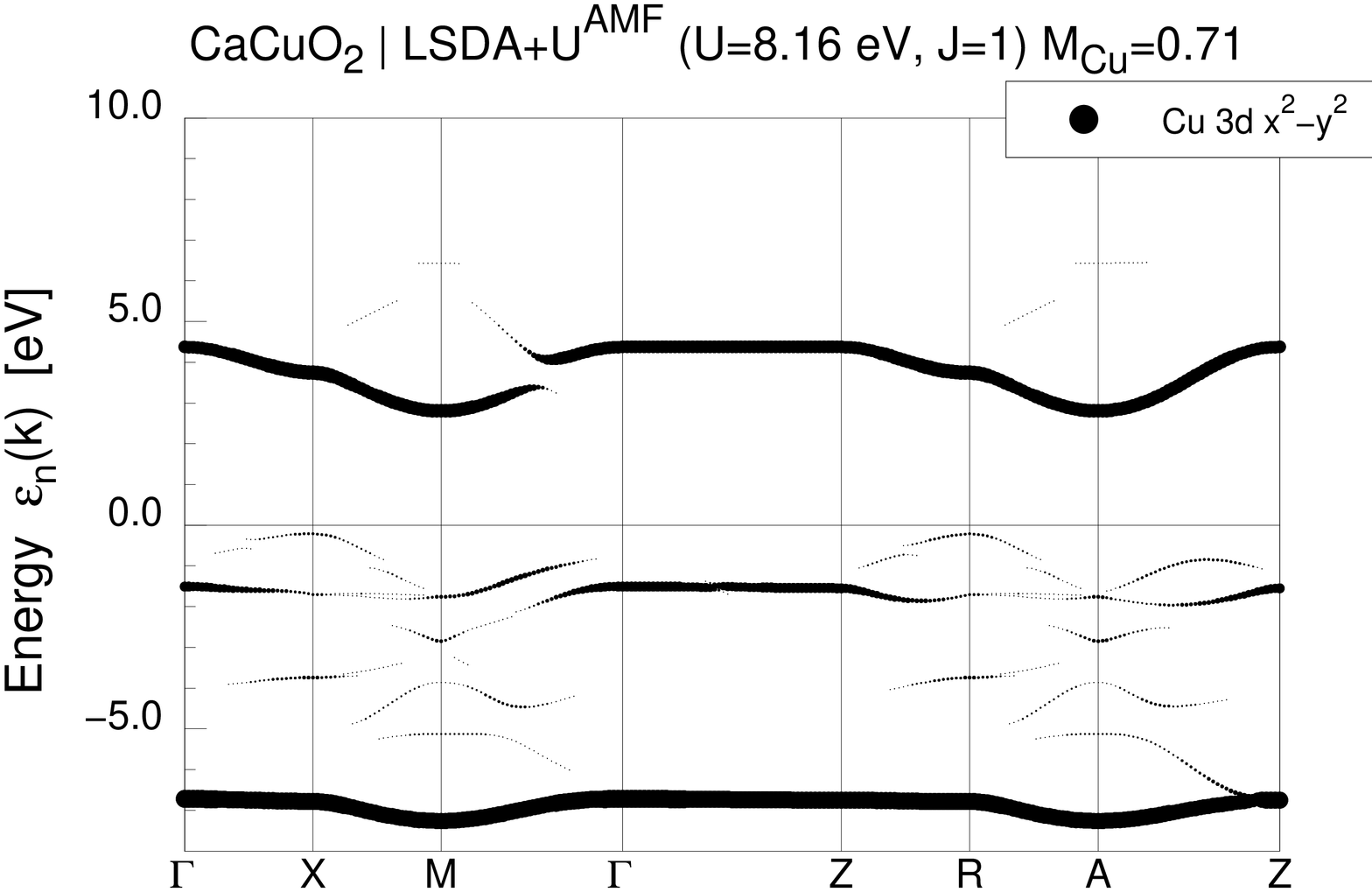}
    \vspace{-1cm}

    \includegraphics[scale=.27,angle=-90]{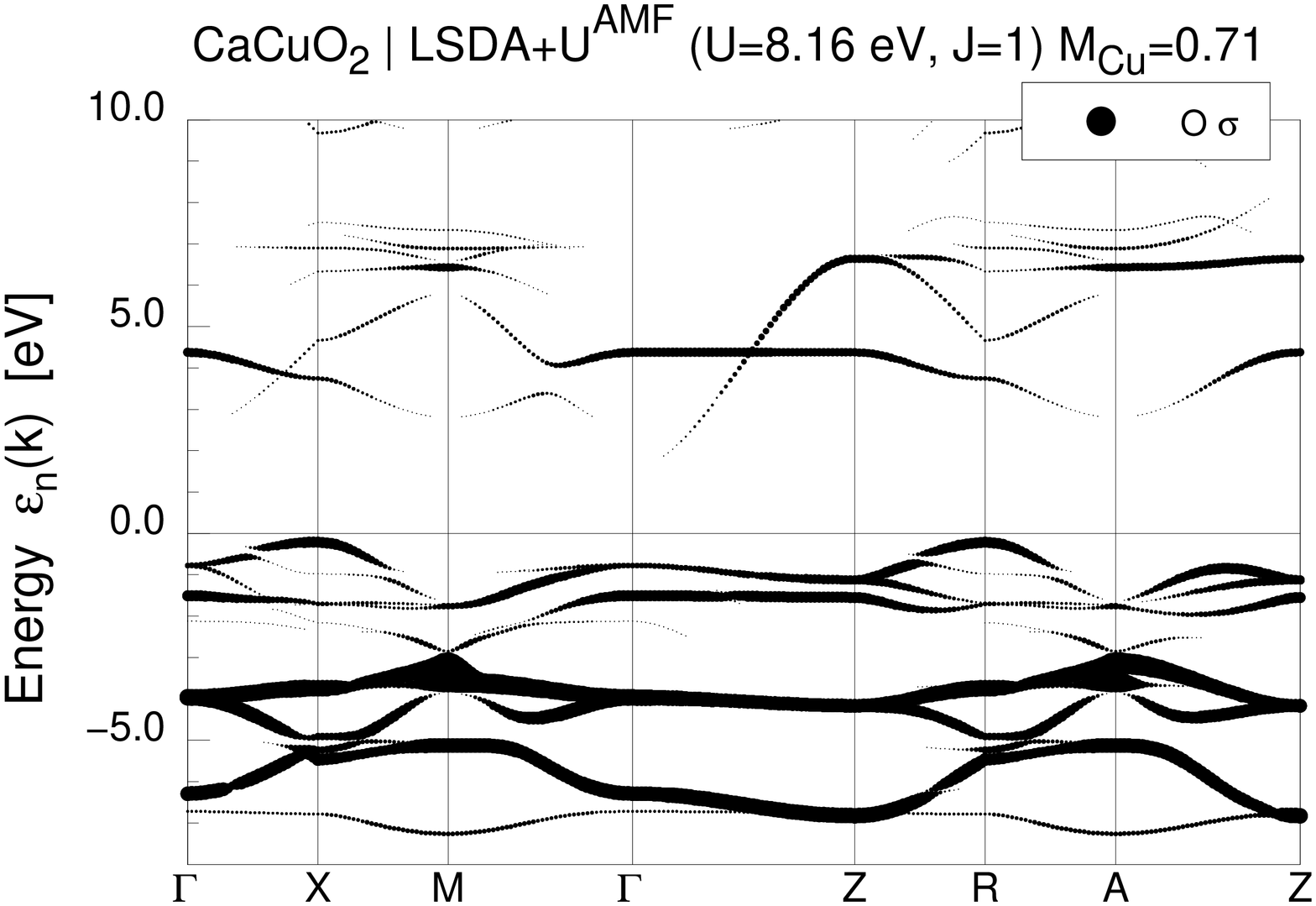}
    \vspace{-1cm}

    \includegraphics[scale=.27,angle=-90]{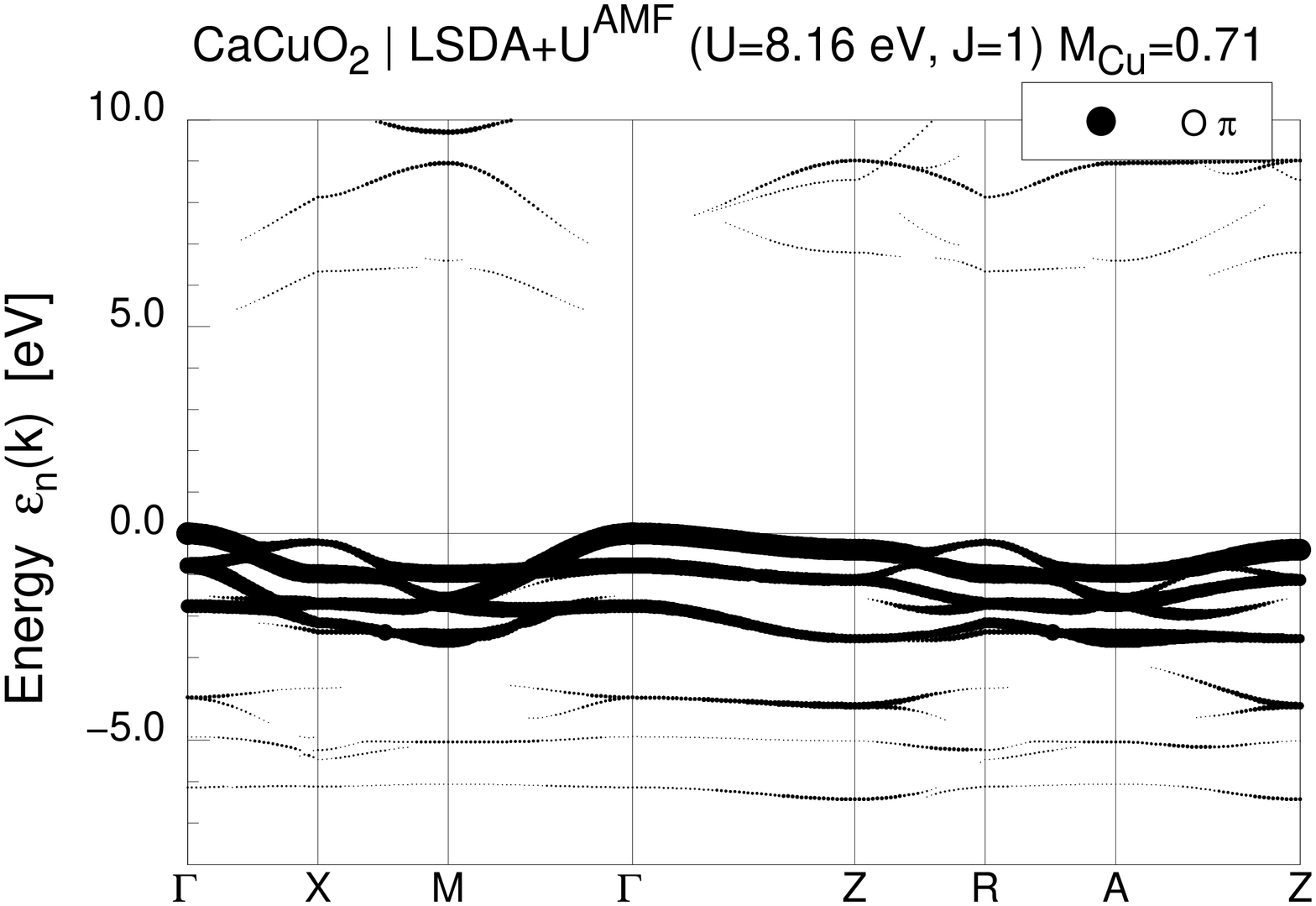}
    \vspace{-1cm}

    \includegraphics[scale=.27,angle=-90]{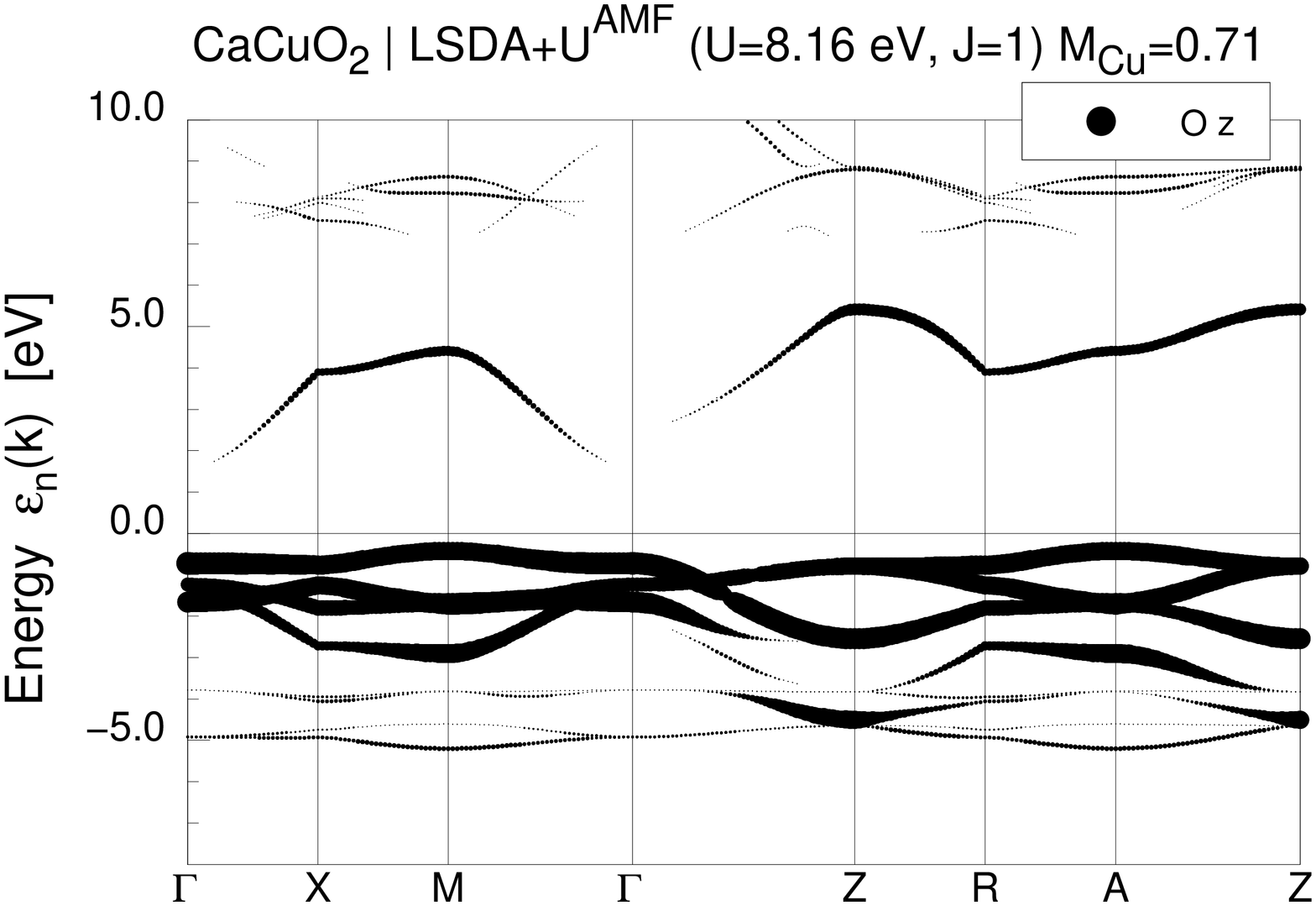}

    \caption{From top to bottom: orbital weight of the Cu-$3d_{x^2-y^2}$
      orbital, the O-$2p_{\sigma}$ orbitals, the O-$2p_{\pi}$ and the O-$2p_z$
      orbitals.} 
    \label{fig5}
  \end{center}
\end{figure}

\begin{figure}
  \begin{center}\vspace*{-2mm}

    \includegraphics[scale=0.45]{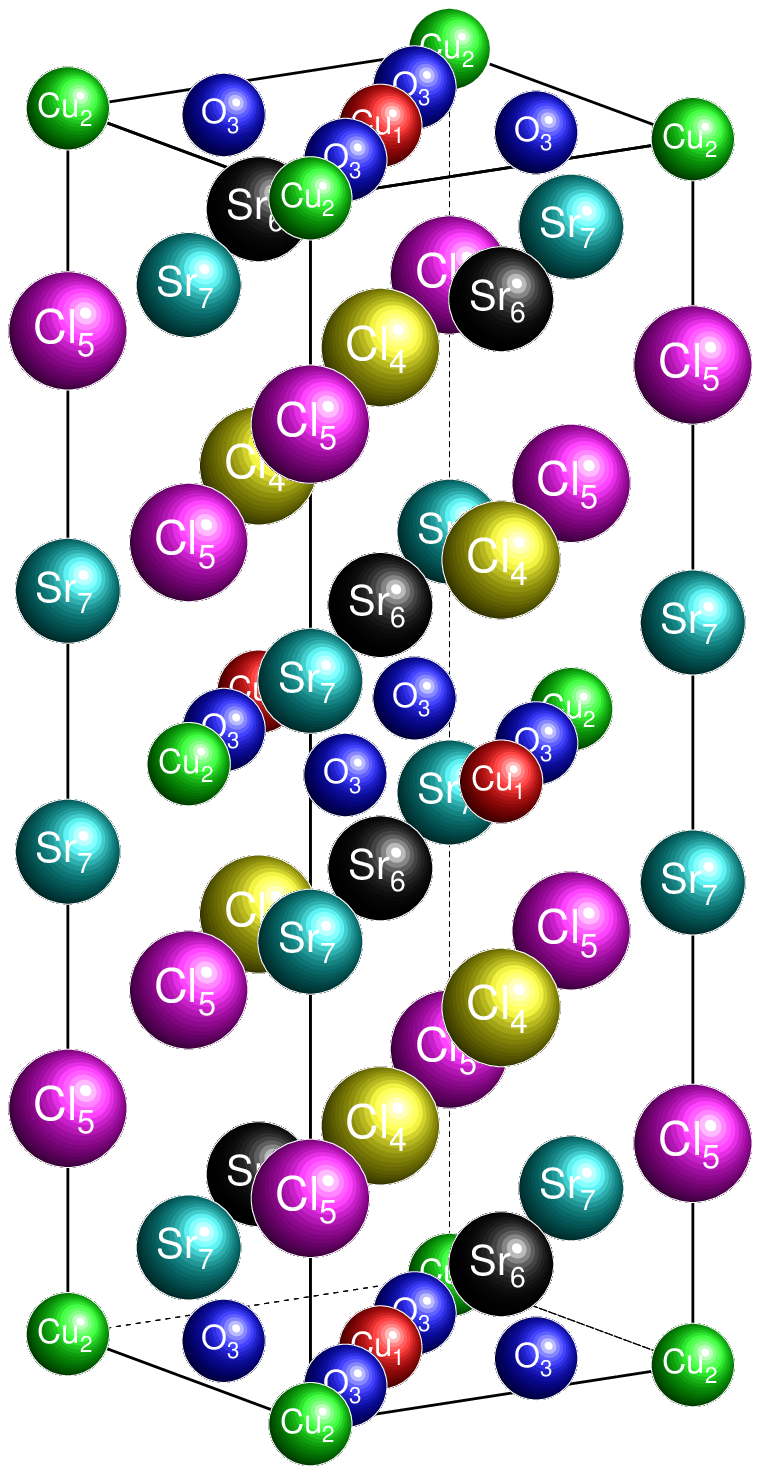}
    \hspace{7mm}
    \includegraphics[scale=0.45]{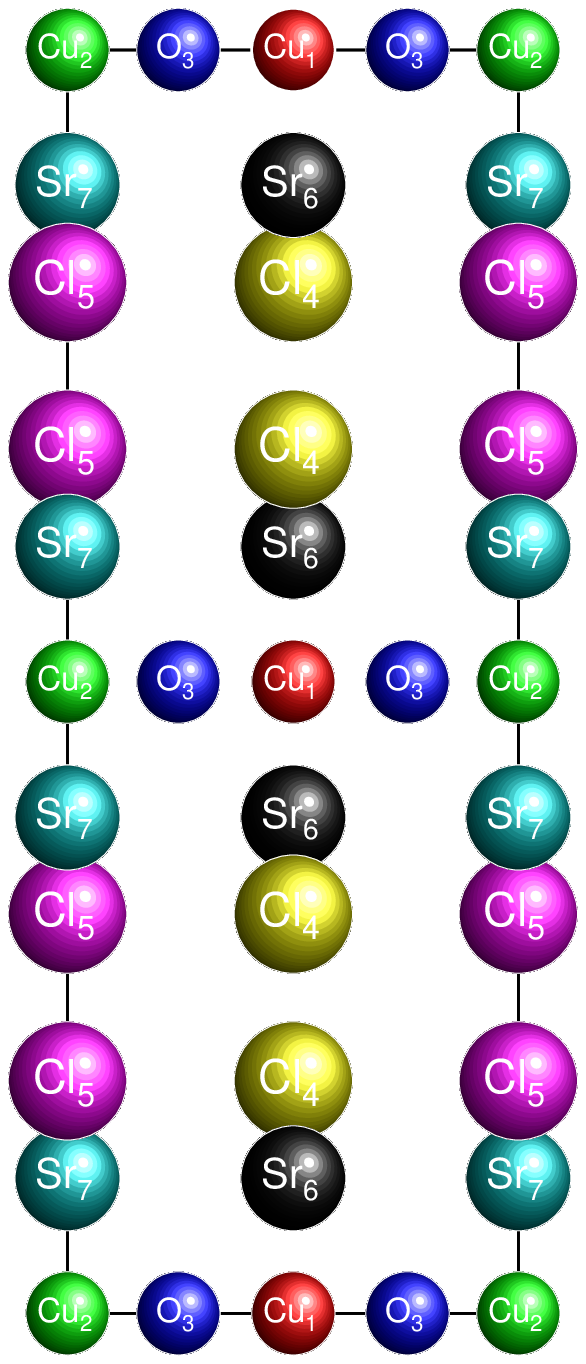}

    \caption{Unit cell of antiferromagnetic Sr$_2$CuO$_2$Cl$_2$; Cu$_1$ spin
      up, Cu$_2$ spin down.}
    \label{fig6}
  \end{center}
\end{figure}

\begin{figure}
  \begin{center}\vspace*{-1cm}
    \includegraphics[scale=.27,angle=-90]{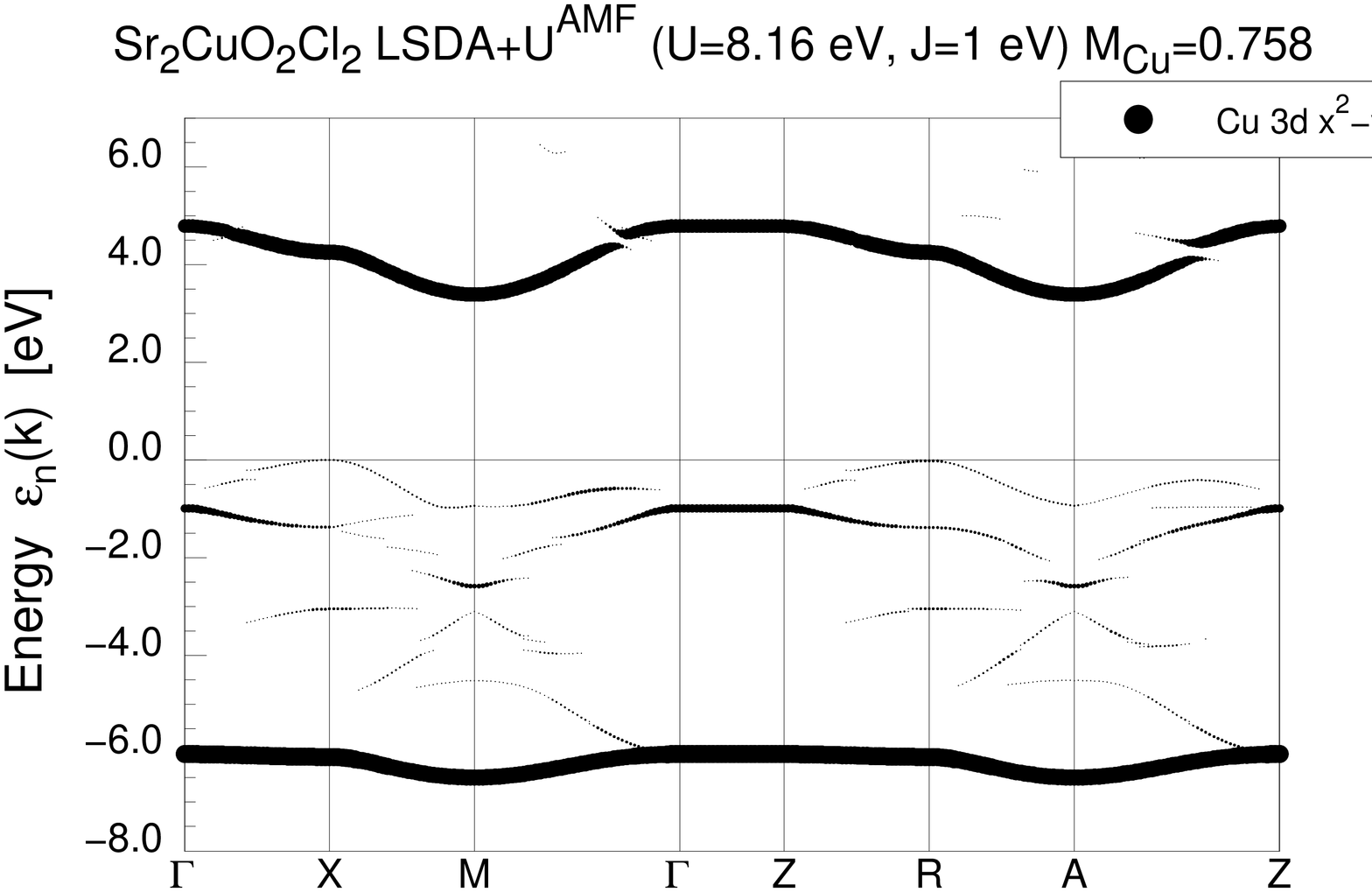}
    \vspace{-1cm}

    \includegraphics[scale=.27,angle=-90]{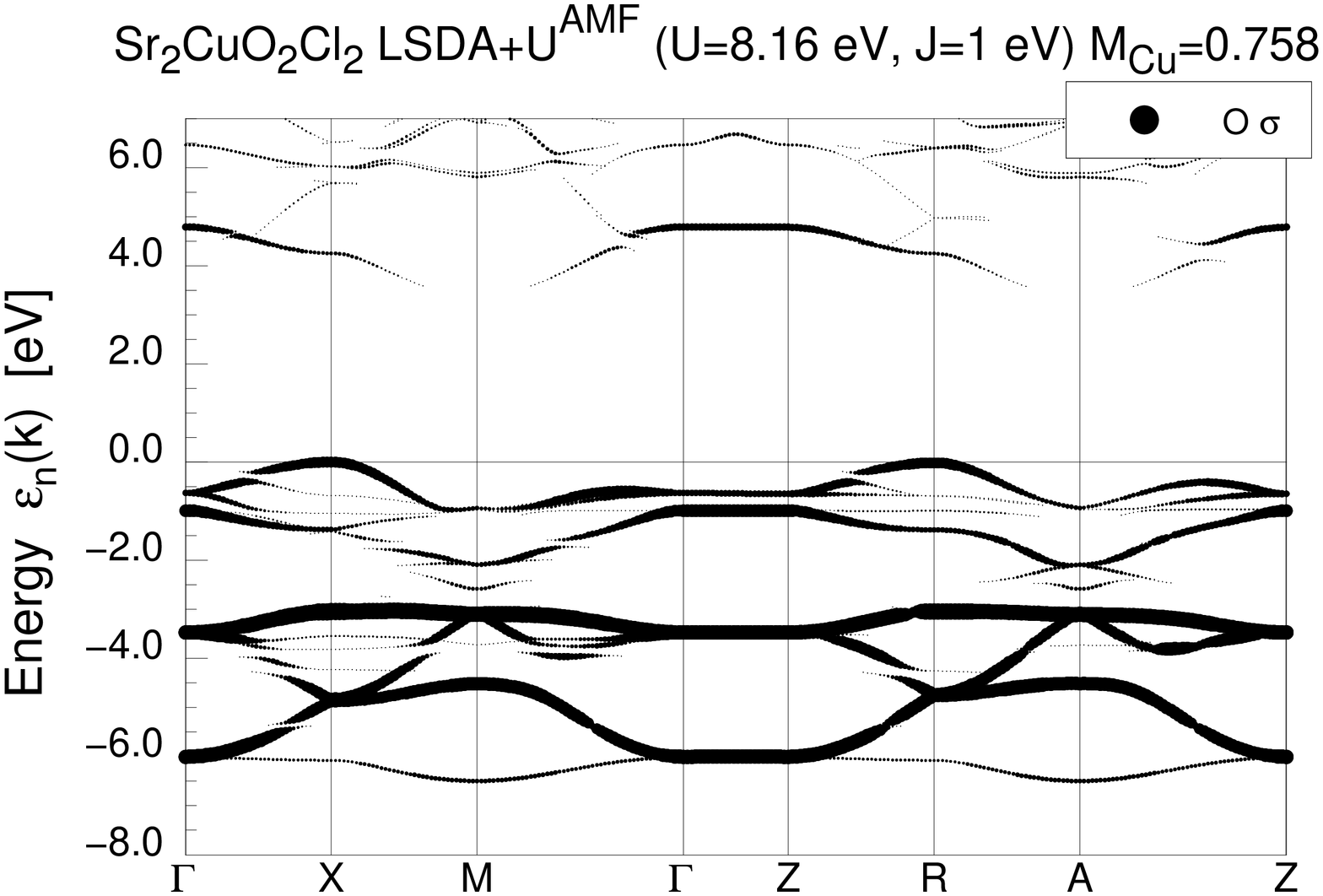}
    \vspace{-1cm}

    \includegraphics[scale=.27,angle=-90]{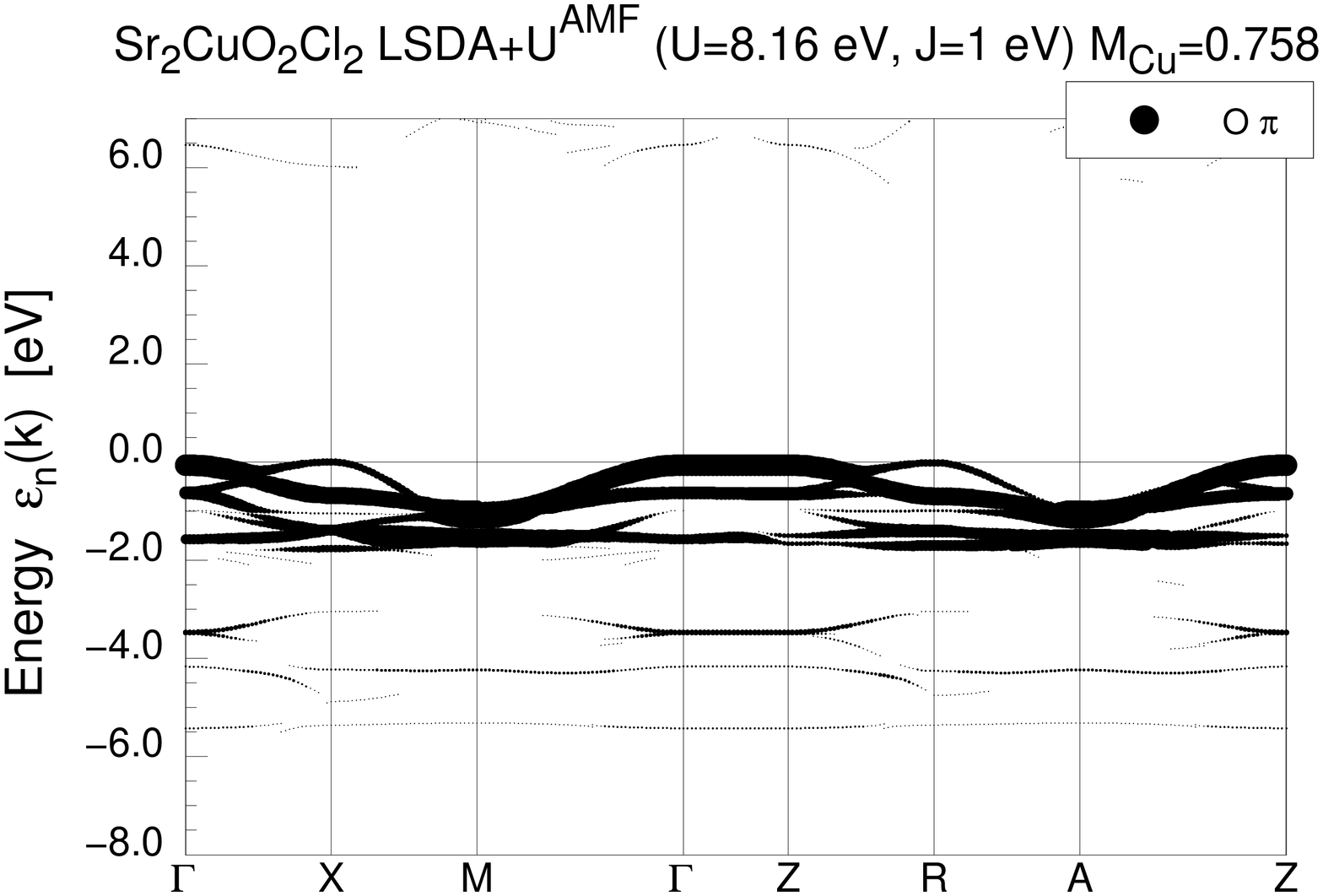}
    \vspace{-1cm}

    \includegraphics[scale=.27,angle=-90]{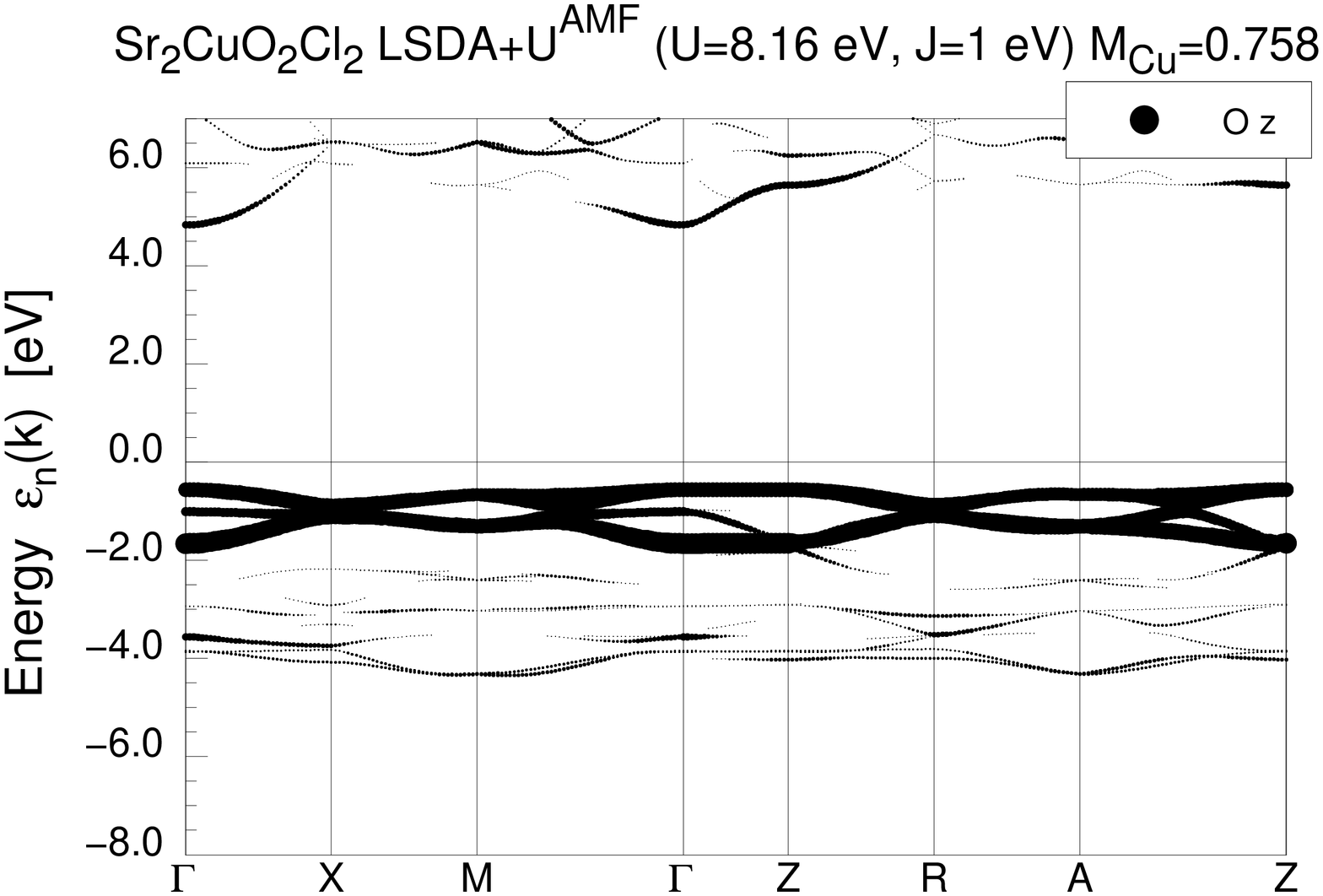}

    \caption{Same as Fig.~\ref{fig5} for Sr$_2$CuO$_2$Cl$_2$.}
    \label{fig7}
  \end{center}
\end{figure}

\begin{figure}
  \begin{center}
    \includegraphics[scale=0.6]{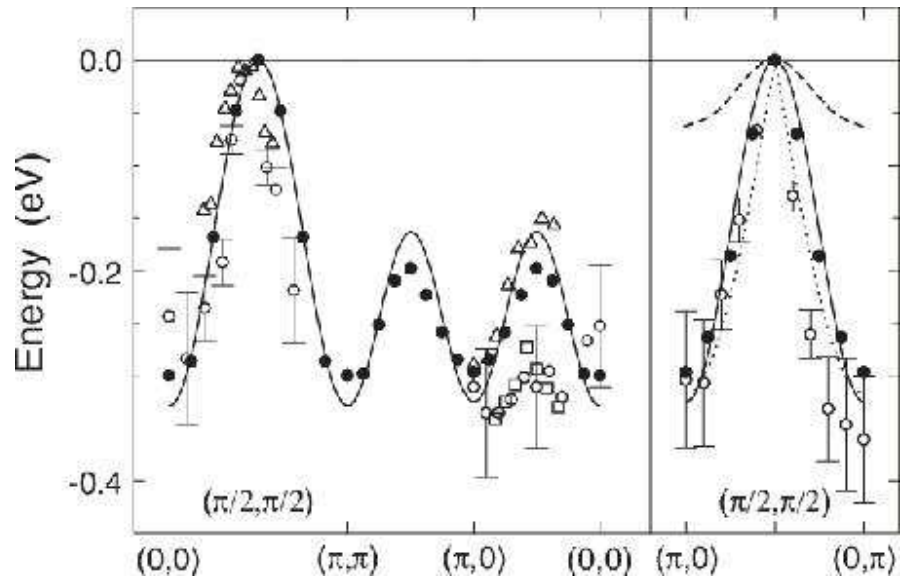}
    \caption{Energy dispersion of quasiparticles for Sr$_2$CuO$_2$Cl$_2$
      \cite{Toh00}. The energy zero is put at the top of the band, about
      0.7 eV below Fermi level. Open symbols: experimental data; solid
      circles: self-consistent Born approximation for a $t-t'-t''-J$
      model; solid line: tight-binding fit; dashed: $t-J$ model;
      dotted: spinon model dispersion.}
    \label{fig9}
  \end{center}
\end{figure}

\begin{figure}
  \begin{center}\vspace*{-1cm}

    \includegraphics[scale=.27,angle=-90]{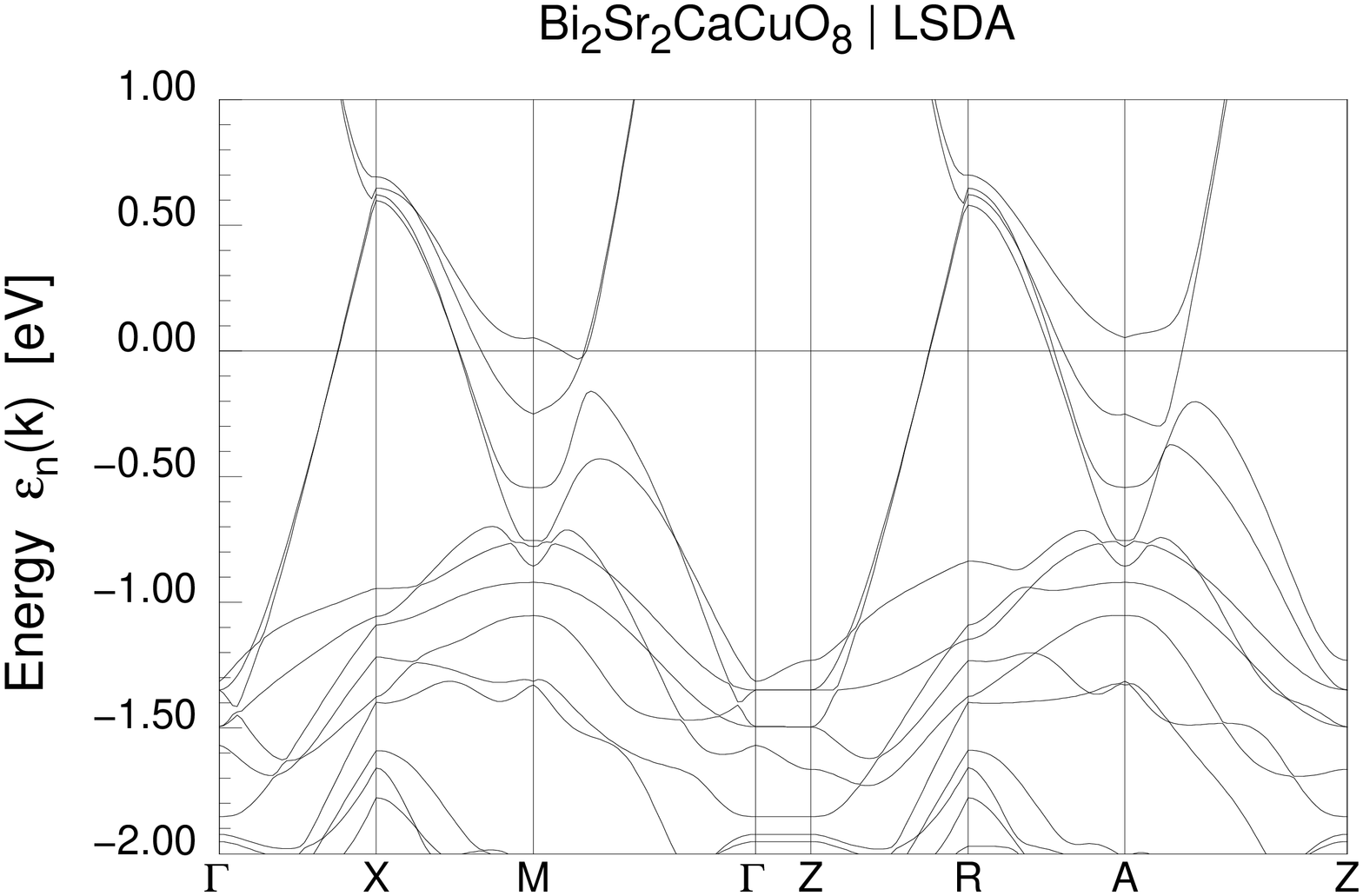}
    \hspace{-1cm}
    \includegraphics[scale=.27,angle=-90]{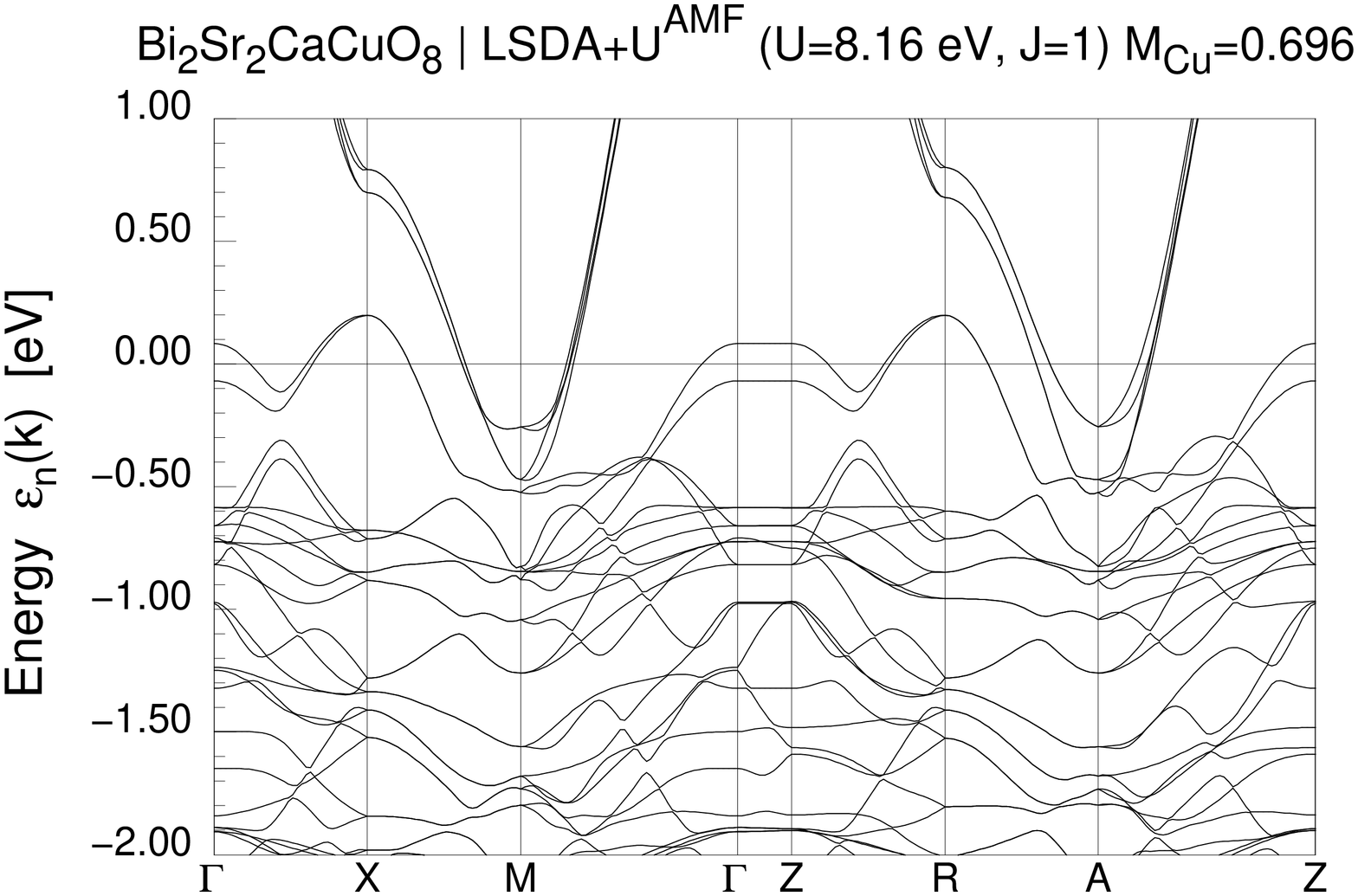}
    \vspace{-1cm}

    \includegraphics[scale=.27,angle=-90]{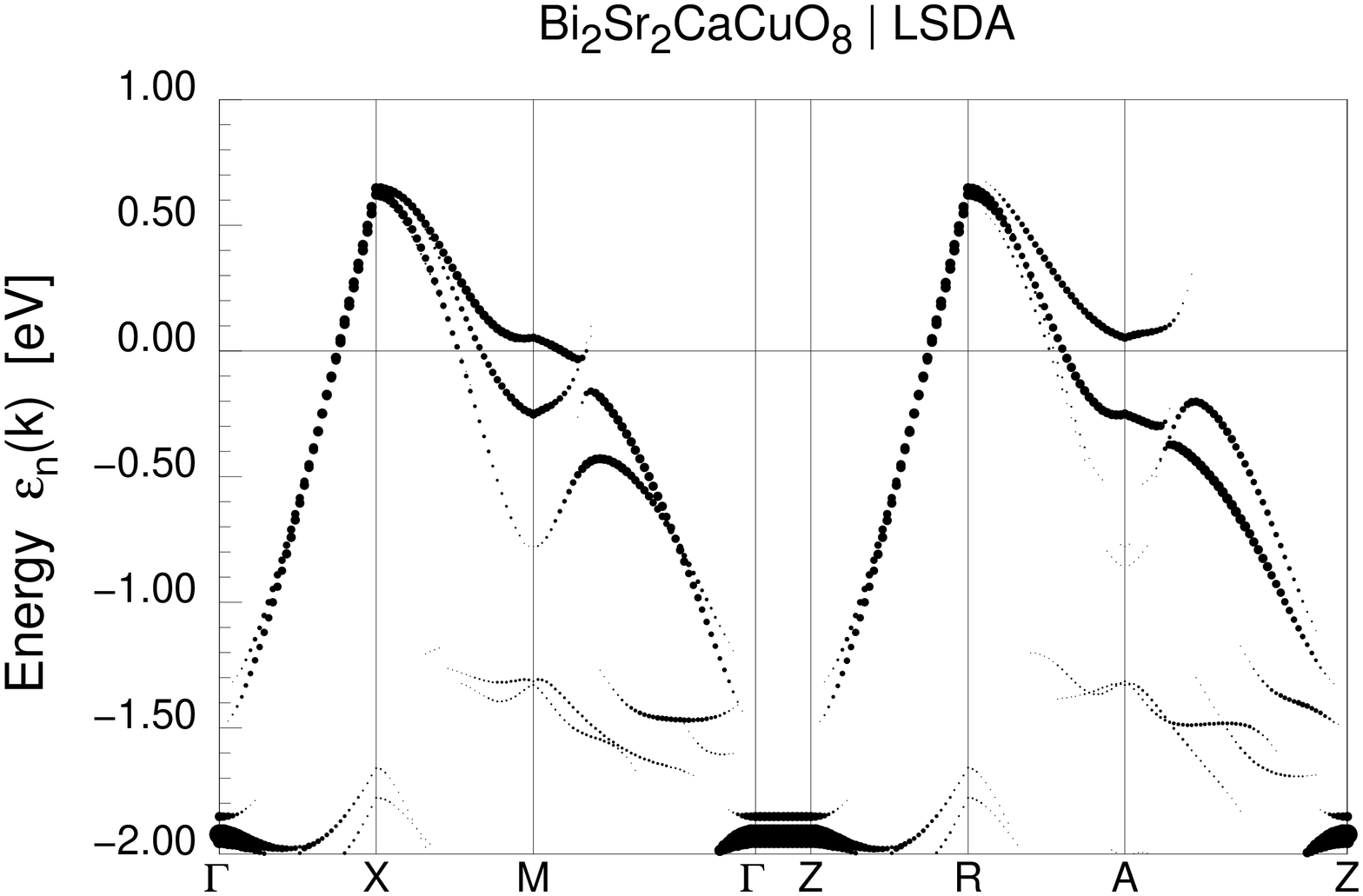}
    \hspace{-1cm}
    \includegraphics[scale=.27,angle=-90]{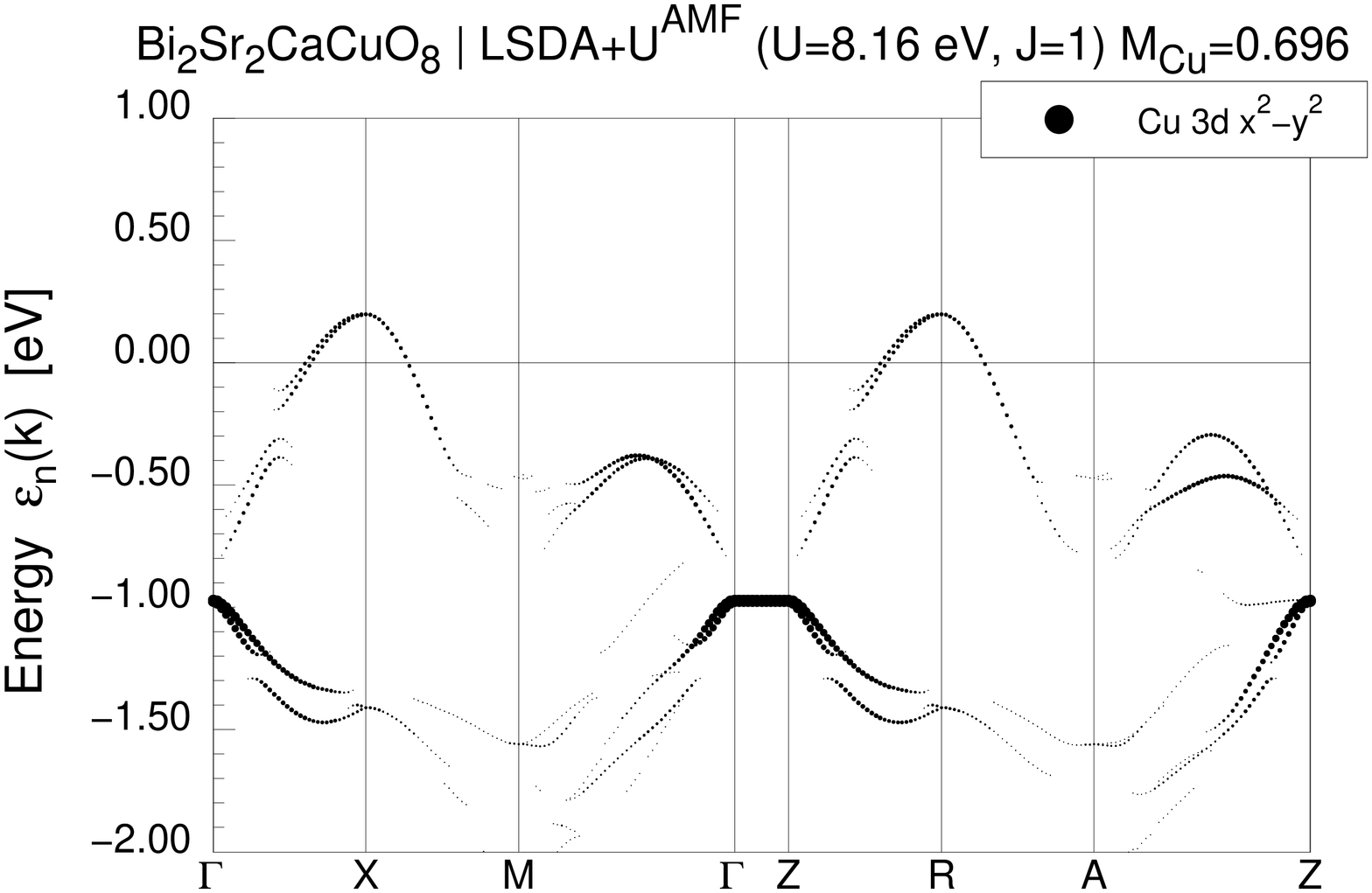}

    \caption{Bi$_2$Sr$_2$CaCu$_2$O$_8$: left: LSDA band structure and
      Cu-$3d_{x^2-y^2}$ orbital weights,
      right: LSDA+$U^\mathrm{AMF}$ bandstructure and weights.}
    \label{figBi}
  \end{center}
\end{figure}

\begin{figure}
  \begin{center}\vspace*{-5mm}
    \includegraphics[scale=.27,angle=-90]{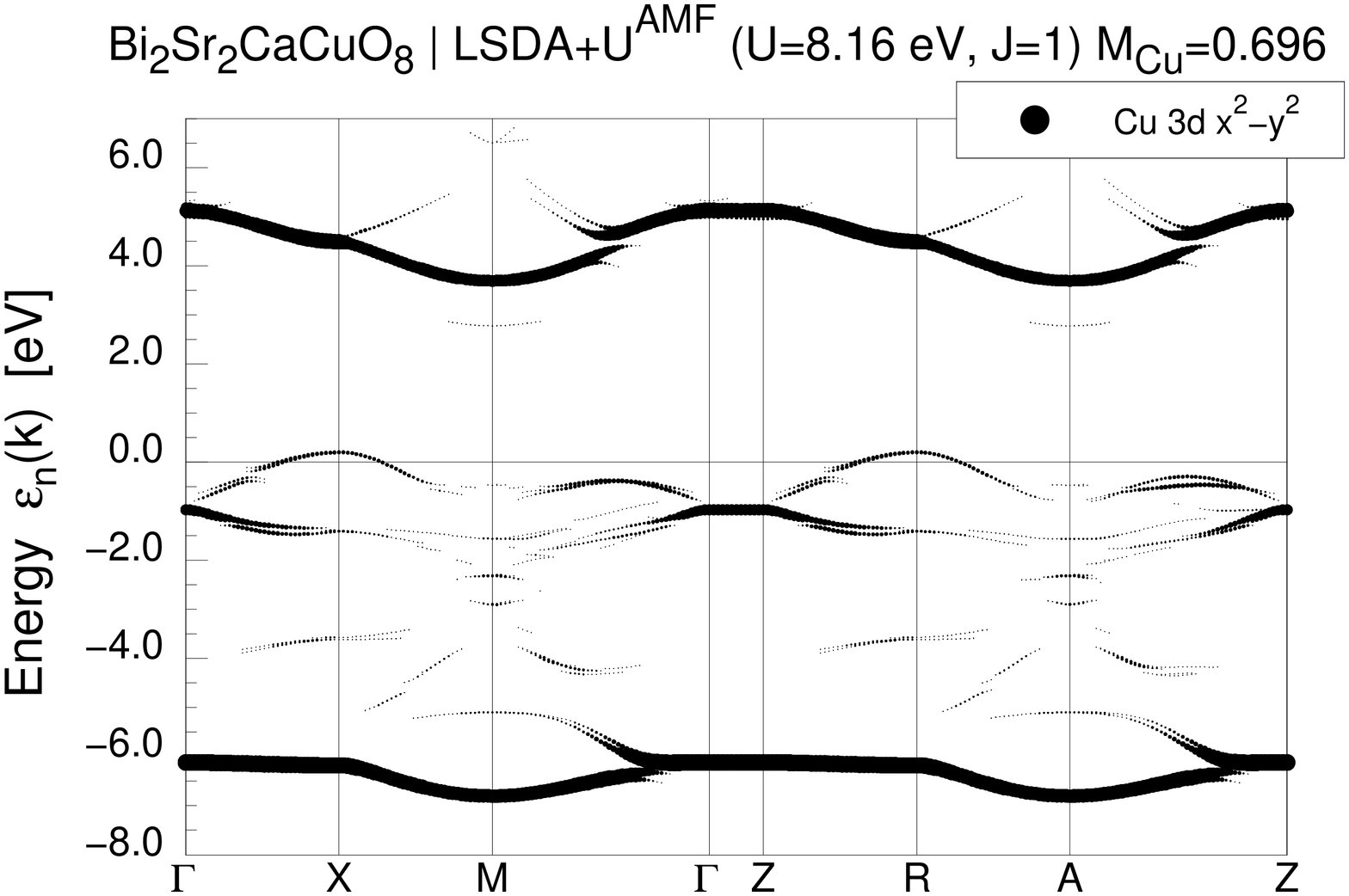}

    \caption{Bi$_2$Sr$_2$CaCu$_2$O$_8$: LSDA+$U^\mathrm{AMF}$
      Cu-$3d_{x^2-y^2}$ orbital weights.}
    \label{figBiCuBig}
  \end{center}
\end{figure}


\begin{thebibliography}{99}

  
\bibitem{Hoh64} P. Hohenberg and W. Kohn, Phys. Rev. 136, B864--71 (1964);
  W. Kohn and L. J. Sham, Phys. Rev. 140, A1133--A1138 (1965).

\bibitem{Lie83} E. H. Lieb, Int. J. Quant. Chem. XXIV, 243--77 (1983).
  
\bibitem{Esc96} H. Eschrig ``The Fundamentals of Density Functional Theory'',
  B. G. Teubner, Stuttgart, 1996; a revised and extended issue is in
  preparation by Stiftung Benedictus Gotthelf Teubner, Leipzig.
  
\bibitem{Her78} J. F. Herbst, R. E. Watson and J. W. Wilkins, Phys. Rev. B17,
  3089--98 (1978).
  
\bibitem{Ded84} P. H. Dederichs \textit{et al.},
  Phys. Rev. Lett. 53, 2512--15 (1984).

\bibitem{Nor86} M. R. Norman and A. J. Freeman, Phys. Rev. B33, 8896--98
  (1986). 

\bibitem{Gun89} O. Gunnarsson \textit{et al.},
  Phys. Rev. B39, 1708--22 (1989).

\bibitem{Pic84} W. E. Pickett and C. S. Wang, Phys. Rev. B30, 4719--33
  (1984). 

\bibitem{Hyb85} M. S. Hybertsen and S. G. Louie, Phys. Rev. Lett. 55,
  1418--21 (1985).

\bibitem{Ani91} V. I. Anisimov, J. Zaanen and O. K. Andersen, Phys. Rev. B44,
  943--54 (1991).

\bibitem{Ani93} V. I. Anisimov \textit{et al.}, 
  Phys. Rev. B48, 16 929--34 (1993).
  
\bibitem{Czy94} M.~T.~Czy\.zyk and G.~A.~Sawatzky, Phys. Rev. B49, 14211--28
  (1994).

\bibitem{Ani97} V. I. Anisimov, F. Aryasetiawan and A. I. Lichtenstein,
  J. Phys.: Condens. Matter 9, 767--808 (1997).
  
\bibitem{Shi99} A.~B.~Shick, A.~I.~Liechtenstein and W.~E.~Pickett, Phys. Rev.
  B60, 10763--69 (1999).
  
\bibitem{Koe99} K.~Koepernik and H.~Eschrig, Phys. Rev. B59, 1743--57
  (1999).
  
\bibitem{Lie95} A.~I.~Liechtenstein, V.~I.~Anisimov and J.~Zaanen, Phys. Rev.
  B52, R5467--70 (1995).

\bibitem{fplo} Access to FPLO exists under `http://www.ifw-dresden.de/fplo'.

\bibitem{Per81} P.~Perdew and A.~Zunger, Phys. Rev. B23, 5048--79 (1981).

\bibitem{Vak89} D. Vaknin \textit{et al.}, 
  Phys. Rev. B39, 9122--25 (1989).
  
\bibitem{Eme87} V.~J.~Emery, Phys. Rev. Lett. 58, 2794--97 (1987).

\bibitem{Zha88} F.~C.~Zhang and T.~M.~Rice, Phys. Rev. B37, 3759--61 (1988).
  
\bibitem{Pot97} J.~J.~M. Pothuizen \textit{et al.}, 
  Phys. Rev. Lett. 78, 717--19 (1997).
  
\bibitem{Hay99} R. Hayn \textit{et al.}, 
  Phys. Rev. B60, 645--58 (1999).

\bibitem{Fin94} J. Fink \textit{et al.}, 
  J. Electron Spectrosc. Relat. Phenom. 66, 395--452 (1994).

\bibitem{Mil90} L.~L.~Miller {\it et al.}, Phys. Rev. B41, 1921--25 (1990).

\bibitem{Toh00} T.~Tohyama, S.~Maekawa, Supercond. Sci. Technol. 13, R17--R32
  (2000).

\bibitem{Dam03} A.~Damascelli, Z-X.~Shen and Z.~Hussain, Rev. Mod. Phys.,
  (2003), in press.

\bibitem{Ron02} F.~Ronning {\it et al.}, cond-mat/0209651 (2002).

\bibitem{Yar02} A.~N.~Yaresko {\it et al.}, Phys. Rev. B65, 115111-1--7
  (2002). 

\bibitem{Poz97} R. Pozzi {\it et al.}, Phys. Rev. B56, 759--65 (1997).

\bibitem{Joh97} D. C. Johnston In K. H. J. Buschow (ed.) ``Handbook of
  Magnetic Materials'', Elsevier, Amsterdam, 1997, Vol. 10,
  Chap. 1. pp. 1--237. 

\bibitem{Gre94} M.~Greven {\it et al.}, Phys. Rev. Lett. 72, 1096--99
  (1994).


\end{thebibliography}
\end{document}